\documentstyle[12pt,epsfig]{article}
\newcommand{\ba}{\begin{array}}
\newcommand{\ea}{\end{array}}
\newcommand{\beq}{\begin{equation}}
\newcommand{\eeq}{\end{equation}}
\newcommand{\bea}{\begin{eqnarray}}
\newcommand{\eea}{\end{eqnarray}}

\def\bce{\begin{center}}
\def\ece{\end{center}}

\def\nonu{\nonumber}

\def\pa{\partial}

\def\be{\beta}

\def\de{\delta}

\def\la{\lambda}

\def\eps6{{\displaystyle \mathop{\epsilon}^{6}}{}}

\def\nab6{{\displaystyle \mathop{\nabla}^{6}}{}}

\def\ba{\begin{array}}
\def\ea{\end{array}}
\def\beq{\begin{equation}}
\def\eeq{\end{equation}}
\def\be{\begin{equation}}
\def\ee{\end{equation}}

\def\la{\lambda}
\def\eps{\epsilon}

\def\ba{\begin{array}}
\def\ea{\end{array}}
\def\beq{\begin{equation}}
\def\eeq{\end{equation}}
\def\be{\begin{equation}}
\def\ee{\end{equation}}

\def\la{\lambda}
\def\eps{\epsilon}

\newcommand{\bean}{\begin{eqnarray*}}
\newcommand{\eean}{\end{eqnarray*}}

\begin{document}
\thispagestyle{empty} \addtocounter{page}{-1}
\begin{flushright}
{\tt hep-th/0504109}\\
\end{flushright}

\vspace*{1.3cm} 
\centerline{\Large
\bf Comments on }
\centerline{ \Large \bf MHV Tree Amplitudes for Conformal
Supergravitons }
\centerline{\Large \bf  
from Topological B-Model 
}
\vspace*{1.5cm}
\centerline{{\bf Changhyun Ahn}
}
\vspace*{1.0cm} 
 \centerline{\it   School of Natural Sciences,
Institute for Advanced Study,
Einstein Drive, Princeton NJ 08540, USA}
\centerline{\it  Department of Physics,
Kyungpook National University, Taegu 702-701, Korea}
\vspace*{0.8cm} 
\centerline{\tt
ahn@ias.edu} 
\vskip2cm

\centerline{\bf Abstract}
\vspace*{0.5cm}

We use the twistor-string theory on the B-model of ${\bf CP}^{3|4}$ 
to compute the maximally helicity violating(MHV) 
tree amplitudes for conformal supergravitons.
The correlator of a bilinear in the affine 
Kac-Moody current(Sugawara stress-energy tensor) 
can generate these amplitudes.
We compare with previous results from open string version
of twistor-string theory.   
We also compute the MHV tree 
amplitudes for both gravitons and gluons from 
the correlators between stress-energy tensor and  current.

\baselineskip=18pt
\newpage
\renewcommand{\theequation}
{\arabic{section}\mbox{.}\arabic{equation}}

\section{Introduction}
\setcounter{equation}{0}

\indent

In \cite{Witten}, the D1-brane instanton contribution in the 
topological B-model of ${\bf CP}^{3|4}$
reproduces the perturbative scattering amplitudes 
in super Yang-Mills theory. 
For tree level MHV amplitudes, D1-brane instanton is a 
complex line ${\bf CP}^1$ 
that is a curve of genus zero and degree one.
On this curve, there is a $(1,0)$ form current 
which is represented by the product of two fermion fields. 
Then the MHV 
scattering amplitude is an integration of the correlators
of the current(motivated by \cite{Nair1988}) 
on D1-brane multiplied by 
external wavefunctions over the holomorphic 
measure(superspace measure)
on the moduli space of D1-brane.
    
In the open string version of twistor-string theory 
\cite{BW}(for the descriptions on open string version, 
see also \cite{Berk,BM}),
the MHV tree amplitudes for conformal supergravitons(in addition 
to gluons) are computed by introducing two different types of
vertex operators of conformal dimension 1 where the basic 
operator product expansion(OPE) behaves a single contraction. 
Each of them describes twistor wavefunction which is 
homogeneous in twistor coordinates of weight 1 and $-1$ 
respectively.  
The scattering amplitude is obtained by 
evaluating the expectation value of the product of the 
vertex operator. The vertex operator consists of 
the above wavefunction and a differential operator of 
bosonic twistor variables. When this differential operator 
acts on the wavefunction containing the exponential,  a single
contraction appears in the amplitude. 
One also sees the inner product between 
the spinors of 
negative helicity when one takes the product of the
wavefunction and the differential operator acting on other 
wavefunction.

It is natural to ask whether the MHV tree amplitudes for
conformal supergravitons can be described in terms of some
correlators on D1-brane multiplied by external wavefunctions
along the line of \cite{Witten,Nair1988}? 
Since the helicity of bottom component of vertex operator 
describing the graviton is equal to 2(recall that the 
helicity of bottom component of vertex operator describing
the gluon is 1), one can think of the quadratic 
expression of current on D1-brane we introduced above as one 
element of the
vertex operator at each external particle. 
That is, so-called Sugawara construction for 
stress-energy tensor.
For the wavefunction, the explicit form 
is already given  
in \cite{BW}. 
There exists also prefactor which depends on
the spinor of positive chirality(parametrizing the ${\bf CP}^1$). 
Then the vertex operator consists of
the wavefunction with a differential operator with respect to 
bosonic twistor variable, correlator for stress-energy tensor
and the prefactor.
We will 
compare our MHV tree amplitudes for 
three and four-point functions
with previous results \cite{BW} from open string version
of twistor-string theory.   
We will 
also compute the MHV amplitudes(three and four-point functions)
for gravitons plus gluons from 
the correlators between stress-energy tensor and the current.

In section 2.1, 
we describe the plane wavefunction of a massless Yang-Mills
particle and introduce a current which has a short-distance behavior.
Then by multiplying each wavefunction with a current, 
the D1-brane 
instanton contribution to $N$-gluons tree level amplitude 
is given. This subsection is a review of \cite{Witten,BW}. 

In section 2.2, 
we consider 
the plane wavefunction of a massless supergraviton
and introduce a stress-energy tensor which is a quadratic 
expression of the current.
The MHV amplitudes for three gravitons and four gravitons
are given and compare our results with \cite{BW}.   

In section 2.3,
we describe the MHV amplitudes including the gluons
for three-point and four-point
from the correlator between  stress-energy tensor and a current.

In section 3,
after discussing the five-point MHV amplitudes briefly,
a generalization for arbitrary $N$-point MHV amplitudes is
considered. Other remarks are given.

For the relevant descriptions for gravitons which are not conformal, 
refer to 
\cite{GRRT,WZ,BBD,Nair2005,BBST,CS,BDI}. 
There are also some works \cite{Polyakov,CGHKM} 
in different context
related to the open string version of 
twistor-string theory \cite{Berk,BM,BW}. 
There are many relevant works on the MHV tree amplitudes and 
we just list some of them here \cite{Refs}. 

\section{MHV tree amplitudes of the B-model of ${\bf CP}^{3|4}$ }
\setcounter{equation}{0}

\indent

\subsection{MHV tree amplitude for gluons }

\indent

In this subsection, we review the results of \cite{Witten,BW}
on MHV tree amplitude for $N$ gluons.
The twistor space wavefunction, corresponding to plane waves 
where translations can be diagonalized, of a massless 
Yang-Mills particle  
with definite momentum $p_{a\dot{a}} = \pi_a 
\widetilde{\pi}_{\dot{a}}$ takes the form \cite{BW}
\bea
\phi(\la,\mu,\psi)=\left(\la/\pi \right)
\de (\langle\la, \pi \rangle) 
\exp \left(
i \left(\pi/\la\right) [\mu,
  \widetilde{\pi}] 
\right)
u \left( \left(\pi/\la\right) \psi \right)
\label{phi}
\eea
which is homogeneous in twistor coordinates $Z^I=(\la,\mu,\psi)$ 
of weight 0
due to the well-defined ratio 
$\pi/\la(=\pi^1/ \la^1)$ and 
$u \left( \left(\pi/ \la\right) \psi \right)$
is the fermionic wavefunction that determines helicity state. 
Under $(\pi,\widetilde{\pi}) \rightarrow (t \pi, t^{-1} 
\widetilde{\pi})$, the $\psi=0$ component of 
$\phi$ scales as $t^{-2}$ implying that this 
operator represents a state in 
Minkowski spacetime of helicity 1(represented by 
a vertex operator that scales as $t^{-2}$ also).
Note that the delta function $\de(\langle\la, \pi \rangle)$
is homogeneous of degree $-1$ in both $\la$ and $\pi$ 
\cite{Witten2004}.

The current $J^r=\alpha T^r \beta$, where $r=1,2, \cdots, 
\mbox{dim} G$ and $T^r$ is a generator in the
fundamental representation of the group,
has a degree of homogeneity in $\pi$ equal to 
$-2$ \cite{Nair1988,BW}.
It is clear that the product of $\phi$ and $J^r$
scales as $t^{-2}$. 
The scattering amplitude obtained by 
calculating the expectation value of the product
of vertex operators will scale as $t^{-2}$ also.
Here 
$\alpha$ and $\beta$ denote
two dimensional free fermions defined on the 
${\bf CP}^1$, that is a curve in 
twistor space, with homogeneous coordinates $\pi_a$. 
The short-distance operator product expansion(OPE) between 
the two currents is given by \cite{KZ,GW}
\bea
J^{r_1}(z_1) J^{r_2}(z_2) & = & 
\frac{k \de^{r_1 r_2}}{z_{12}^2}  
+\frac{  f^{r_1 r_2 r_3} J^{r_3}(z_2)}{z_{12}} + \cdots 
\label{jjope}
\eea
where $k$ is a central charge or level, $f^{r_1 r_2 r_3}$ is a
structure constant, the dots are 
an infinite set of other regular terms and $z_{12} \equiv z_1-z_2$.
The construction of free fermion fields
$\alpha$ and $\beta$ leads to a central charge $k=1$
from their defining OPE's 
$\beta(z_1) \alpha(z_2) =\frac{1}{z_{12}} +\cdots$ 
\cite{Nair1988} by plugging  $J^{r_i}=\alpha T^{r_i} 
\beta$ into (\ref{jjope})
and computing the OPE.  
By acting the vacuum state on the above OPE (\ref{jjope}), 
the two-point function is 
given by 
\bea
\left\langle J^{r_1}(z_1) J^{r_2}(z_2) \right\rangle  & = & 
\frac{k \de^{r_1 r_2}}{z_{12}^2} 
\nonu 
\eea
and three-point function can be obtained from (\ref{jjope}) 
similarly
\bea
\left\langle J^{r_1}(z_1) J^{r_2}(z_2) J^{r_3}(z_3) 
\right\rangle & = & 
\frac{ k f^{r_1 r_2 r_3}}{
z_{12} z_{23} z_{31}}.
\label{jjjcor}
\eea
Moreover, when we compute higher correlation function,
due to the double contractions between the current algebra,
there exist multi-trace contributions in the amplitudes
\footnote{  
For example, one can compute the four-point function explicitly
$
\left\langle  J^{r_1}(z_1) J^{r_2}(z_2) J^{r_3}(z_3) J^{r_4}(z_4) 
\right\rangle  =  k^2 \left( \frac{\de^{r_1r_2} 
\de^{r_3 r_4}}{z_{12}^2 z_{34}^2} 
+ (2 \leftrightarrow 3)
+(2 \leftrightarrow 4)
\right) 
 + k 
\left(
 \frac{f^{r_1 r_2 r_5} f^{r_3 r_4 r_5}}
{z_{12}z_{23} z_{24} z_{34}} 
+ (2 \leftrightarrow 3)
+ (2 \leftrightarrow 4)
\right)$ 
where the symbol $(i \leftrightarrow j)$ 
means that we simply exchange $z_i$ and
$z_j$(and $r_i$ and $r_j$) for 
each explicit quantity. The disconnected part of the 
current correlator(for example,  $k^2 \frac{\de^{r_1r_2} 
\de^{r_3 r_4}}{z_{12}^2 z_{34}^2}$ term) 
leads to  a double contraction in both
$z_1-z_2$ and $z_3-z_4$ channels \cite{Witten}.
By writing $z_{24}=z_{21} +z_{14}$ in the first term of
linear in $k$ above, one can extract the term 
$\frac{1}{z_{12} z_{23} z_{34} 
z_{41}}$ in four-point correlator 
by ignoring the double contraction
of $z_{12}^2$.
\label{foot1} }.

Using the homogeneous coordinates $(\la, \mu, \psi)$ of 
${\bf CP}^{3|4}$ to set $\la_i^1=1$ for all $i$.
A degree 1 instanton has $z_i = \la^2_i$ \footnote{We denote
$(\la_k, \mu_k, \psi_k)=(\la(z_k),\mu(z_k),\psi(z_k))$ and 
parametrize the image of $z_k$ in twistor space.}.
Then there exist following relations \cite{BW}
\bea
\la_k/\pi_k =\la_k^1/ \pi_k^1 =1/ \pi_k^1, 
\qquad  z_{jk} = z_j-z_k= -\frac{\langle \pi_j, \pi_k 
\rangle}{\pi_j^1 \pi_k^1} =
\frac{\langle j, k 
\rangle}{\pi_j^1 \pi_k^1}.
\label{relations}
\eea
In the second relation, we already use the
delta function constraint: 
the vertex operator for the $j^{th}$ external particle
has a delta function $\de (\langle\la_j, \pi_j \rangle)=
\de(\pi_j^2-z_j \pi_j^1)$. Therefore, 
each delta function leads to a factor $(\la_j/\pi_j)$
due to the following integral
\bea
\int d z_j \de (\pi_j^2 -z_j \pi_j^1) = (\la_j/\pi_j). 
\label{deltacon}
\eea   
The left hand side has a homogeneity $2-1=1$ in $\la_j$
because $dz_j=\langle\la_j, d \la_j\rangle$ which has a degree
2 and the delta function has a degree $-1$. This is
consistent with the homogeneity for the right hand side.

By multiplying each wavefunction 
$\phi_j(\la,\mu,\psi)$ (\ref{phi}) characterized by
$j^{th}$ external particle with 
the current $J^{r_j}$ 
and taking the vacuum expectation values,
the D1-brane 
instanton contribution of degree 1 to $N$-gluon tree level 
MHV amplitude is given by \cite{Witten}
\bea
\int d^4 x   d^8 \theta \left(
\prod_{j=1}^N d z_j 
\left(\la_j/\pi_j \right)
\de (\langle\la_j, \pi_j \rangle) 
\exp \left(
i \left(\pi_j/\la_j\right) [\mu,
  \widetilde{\pi}_j] 
\right)
u \left( \left(\pi_j/\la_j\right) \psi_j \right) \right)
\left\langle \prod_{k=1}^{N} J^{r_k}(z_k) \right\rangle
\nonu
\eea
where the $N$-point current correlator for single-trace
amplitude
can be generalized and
written in terms of $\la_i$ and $\pi_i$ and 
is given by 
\bea
\left\langle \prod_{k=1}^{N} J^{r_k}(z_k)\right\rangle =
\prod_{i=1}^{N} \frac{1}{z_{i (i+1)}} =
\prod_{i=1}^{N}
\frac{\left(\pi_i/ \la_i \right)^2}{
\left\langle i, i+1\right\rangle}
\label{currentcor}
\eea 
with a group theory factor
and this is homogeneous of degree 0 in each 
$\pi_i$(recall that $i=\pi_i$)
and $-2$ in each $\la_i$.
We used the second relation of (\ref{relations}).
Now it is easy to see the factor 
$\left(\la_j/ \pi_j \right)^2$ coming 
from both the wavefunction and a delta function is canceled 
out the  factor $\left(\pi_j/ \la_j \right)^2$
in the current correlator (\ref{currentcor}).
In the integral over $x_{a\dot{a}}$, 
we use the twistor equation 
$\mu^{\dot{a}} = x^{a\dot{a}} \la_a$ and this will lead to
the usual delta function of energy-momentum conservation
and the integral over the $\theta^{Aa}$ will give us a
factor $\langle r, s\rangle^4$ where the particles $r$ and $s$
have negative helicity in their multiplets.

The final expression for MHV amplitude for $N$ gluons 
in terms of correlation functions of 
currents on ${\bf CP}^1$ is then  
\bea
(2\pi)^4 \de^4 \left(\sum_{j=1}^{N} p_{j}^{a \dot{a}} \right) 
\langle r, s\rangle^4
\prod_{i=1}^{N} \frac{1}{\left\langle i, i+1\right\rangle}.
\nonu
\eea
In next subsections, we would like to see 
the MHV tree
amplitude for $N$ external particles including the
supergravitons and gluons.

\subsection{MHV tree amplitude for supergravitons }

\indent

The twistor space wavefunction corresponding to plane waves 
of a massless 
supergraviton with definite momentum $p_{a\dot{a}} = \pi_a 
\widetilde{\pi}_{\dot{a}}$ is given by \cite{BW}
\bea
f^{\dot{a}}(\la,\mu,\psi) = 
\left(\la/ \pi \right)^2
\widetilde{\pi}^{\dot{a}} 
\de (\langle\la, \pi \rangle) 
\exp \left(
i \left(\pi/ \la\right) [\mu,
  \widetilde{\pi}] 
\right)
u \left( \left(\pi/ \la\right) \psi \right)
\label{f}
\eea
which is homogeneous in twistor coordinates $Z^I=(\la,\mu,\psi)$ 
of weight 1.
Under $(\pi,\widetilde{\pi}) \rightarrow (t \pi, t^{-1} 
\widetilde{\pi})$, the $\psi=0$ component of 
$f^{\dot{a}}$ scales as $t^{-4}$ implying that this 
operator represents a state in Minkowski spacetime of helicity 2.

What is vertex operator for graviton?
One can think of quadratic expression of  
the current, $J^r J^r(z)$,
which is nothing but stress-energy tensor $T(z)$, 
has a degree of homogeneity in $\pi$ equal to 
$-4$ because $J^r(z)$ has a degree of homogeneity equal to 
$-2$.
Obviously, the product of $f^{\dot{a}}$ and $J^r J^r$
scales as $t^{-4}$. As for gluons, 
two dimensional free fermions, defined on the 
${\bf CP}^1$ with homogeneous coordinates $\pi_a$, can be
constructed. 
The exact coefficient of $J^r J^r$ in $T(z)$
depends on both the level of affine Kac-Moody algebra and 
dual Coxeter number of underlying finite dimensional Lie algebra.

Among possible vector field $f^I \pa_I$,
only on the vector fields $f^{\dot{a}} \pa_{\dot{a}}$, 
the translation generators $c^{a\dot{a}} \la_a 
\frac{\pa}{\pa \mu^{\dot{a}}}$, where $c^{a\dot{a}}$ 
is a constant vector in the spinor notation, 
can be diagonalized. 
Recall that in \cite{BW}, the external particles have the 
wavefunctions which are plane waves and in this case, translations
can be diagonalized. We also restrict ourselves to this particular
wavefunctions. 

We expect that the scattering amplitudes can be obtained 
by multiplying each wavefunction(vector field) with 
stress-energy tensor and taking the vacuum expectation values. 
Let us describe each case in detail.

$\bullet$ Three-point amplitude for three gravitons

In order to compute the amplitude, one has to 
get an expression for the correlators between stress-energy
tensor.
It is very well-known that
the short-distance OPE 
between the two stress-energy tensors in two dimensional 
conformal field theory \cite{BPZ,Ginsparg} is given by
\bea
T(z_1) T(z_2) & = & \frac{c/2}{z_{12}^4} + \frac{2 T(z_2)}{z_{12}^2} 
+\frac{\partial T(z_2)}{z_{12}} + \cdots 
\label{ttope}
\eea
where $c$ is a central charge and the dots stand for the terms 
regular in the limit $z_1 \rightarrow z_2$ and we use a simplified 
notation $z_{12} \equiv z_1-z_2$.
The vacuum expectation value of this OPE(two-point correlation
function) is  written as
\bea
\langle T(z_1) T(z_2) \rangle & = & \frac{c/2}{z_{12}^4}. 
\nonu  
\eea
One can also see this fact 
from the mode expansion for $T(z)$ by requiring
the regularity of $T(z)|0\rangle$ at $z=0$ \cite{Ginsparg}.
This two-point function is invariant under translations, 
rotations, dilatations and special conformal transformation.
The three-point correlation function
can be obtained from the above OPE (\ref{ttope}):
\bea
\langle T(z_1) T(z_2) T(z_3) \rangle 
& = & \frac{c}{z_{12}^2 z_{23}^2 z_{31}^2}.
\label{tttcor}
\eea
The three points $z_1, z_2$, and $z_3$ can always be mapped 
to three reference points, say, $\infty, 1$, and 0 by a conformal
transformation. This is what we need in order to compute the 
three particle scattering amplitude.  
This looks like (\ref{jjjcor}), but 
the double contractions are present.

With this preliminary view,
let us, first, consider the subamplitude where
each $\mu$-derivative $\frac{\pa}{\pa \mu_{i+1}}$ 
of vertex operator at each 
$i^{th}$ particle acts on next $(i+1)^{th}$ particle.
The adjoining two spinor indices between
$\widetilde{\pi}_i^{\dot{a}}$ from the wavefunction (\ref{f}) and 
$\mu_{i+1}^{\dot{a}}$ in the $\mu$-derivative 
are contracted each other.
Remember that in the notation $\mu_i^{\dot{a}}$,
the subscript $i$ implies the $i^{th}$ particle the $\mu$-derivative
acts on and the spinor index $\dot{a}$ implies the vertex 
operator it belongs to. 
Since 
the $\mu$'s of the twistor variable 
in the wavefunctions (\ref{f}) appear only in the exponentials,
one can put each $\mu$-derivative 
$\frac{\pa}{\pa \mu_{i+1}}$ 
in front of 
corresponding each $(i+1)^{th}$ exponential.
We introduce a prefactor 
$\left(\frac{\pi_i^a \pi_{ib}}{\pi_i^1 \pi_i^1}\right)$, 
which is 
obviously homogeneous of degree 0 in $\pi_i$(does not spoil
the behavior of vertex operator by scales under $\pi$ and 
$\widetilde{\pi}$) and degree 2 
in $\la_i$ by recalling (\ref{relations}),
in the vertex operator at the $i^{th}$ particle.  
Each spinor index at $i^{th}$ particle in this prefactor
is contracted with both
the one at $(i+1)^{th}$ particle and the one at $(i-1)^{th}$
particle.
The role of this factor, on the one hand, 
makes the scattering amplitude to be homogeneous
of degree 0 in each $\la_i$.
Even if the individual degree of each $\la_i$
does not vanish, one can find the appropriate prefactors
make the sum of degrees of $\la_i$'s vanish.
We will see this in next example.  
On the other hand, it kills the factor $z_{i(i+1)}$
in the correlator (\ref{tttcor}).

The vertex operator-integrand for this particular case is 
given by the prefactors, correlator, and 
the wavefunctions: $
  \left(\frac{\pi_1^{a} \pi_{1c}}{\pi_1^1 \pi_1^1} \right)
  \left(\frac{\pi_2^{b} \pi_{2a}}{\pi_2^1 \pi_2^1} \right) 
   \left(\frac{\pi_3^{c} \pi_{3b}}{\pi_3^1 \pi_3^1}\right)
\left\langle  
\prod_{k=1}^3 T(z_k)  
\right\rangle
f^{\dot{a}}_1 \frac{\pa}{\pa \mu^{\dot{a}}_2}
f^{\dot{b}}_2 \frac{\pa}{\pa \mu^{\dot{b}}_3}
f^{\dot{c}}_3 \frac{\pa}{\pa \mu^{\dot{c}}_1}
$.
Let us write this vertex operator explicitly 
in order to compute the subamplitude as follows:
\bea
\int d^4 x d^8 \theta dz_1 dz_2 dz_3 
&&   \left(\frac{\pi_1^{a} \pi_{1c}}{\pi_1^1 \pi_1^1} \right)
\de (\langle\la_1, \pi_1\rangle) 
\left(\la_1/\pi_1 \right)^2
\widetilde{\pi}_1^{\dot{a}} \frac{\pa}{\pa \mu_2^{\dot{a}}} 
\exp \left(
i (\pi_2/\la_2) [\mu_2,
  \widetilde{\pi}_2] 
\right)
 \nonu \\
&&   \left(\frac{\pi_2^{b} \pi_{2a}}{\pi_2^1 \pi_2^1} \right) 
\de (\langle\la_2, \pi_2\rangle) 
\left(\la_2/\pi_2 \right)^2 
\widetilde{\pi}_2^{\dot{b}} \frac{\pa}{\pa \mu_3^{\dot{b}}} 
\exp \left(
i (\pi_3/\la_3) [\mu_3,
  \widetilde{\pi}_3] 
\right)
 \nonu \\
&&   \left(\frac{\pi_3^{c} \pi_{3b}}{\pi_3^1 \pi_3^1}\right) 
\de (\langle\la_3, \pi_3\rangle) 
\left(\la_3/\pi_3 \right)^2 
\widetilde{\pi}_3^{\dot{c}} \frac{\pa}{\pa \mu_1^{\dot{c}}} 
\exp \left(
i (\pi_1/\la_1) [\mu_1,
  \widetilde{\pi}_1] 
\right)
 \nonu \\
&& 
\left\langle  
\prod_{k=1}^3 T(z_k) 
\right\rangle
 \prod_{j=1}^3 u \left((\pi_j/\la_j) \psi_j \right). 
\label{122331}
\eea
Let us simplify this by performing $z_i$
integral first.
After differentiating the exponentials 
with respect to each $\mu$'s, 
the factor $[1,2][2,3][3,1]$ for the spinors of
negative chirality arises in this amplitude together 
with a factor
$i^3 \prod_{i=1}^3 (\pi_i/ \la_i)$. 
Using the second relation of (\ref{relations})(in other words, 
by changing the prefactors into $z_{i(i+1)}$'s), 
the factor $z_{12} z_{23} z_{31}$ 
in the denominator of the correlator $\langle  
T(z_1) T(z_2) T(z_3) 
\rangle$, characterized by (\ref{tttcor}), is canceled out. 
As a result,  the remaining term leads to 
the factor $\langle1,2 \rangle \langle2,3 \rangle
\langle 3,1\rangle$ for the spinors of
positive chirality in the denominator of this
amplitude. 
An extra factor $(\la_i/ \pi_i)$ appears by computing the 
delta function integration (\ref{deltacon}).
As for gluons, for the integral over $x_{a\dot{a}}$, 
we use the twistor equation 
$\mu^{\dot{a}} = x^{a\dot{a}} \la_a$ and this will lead to
the usual delta function of energy-momentum conservation
and the integral over the $\theta^{Aa}$ will give us a
factor $\langle r, s\rangle^4$ where particles $r$ and $s$ 
have the minimum helicity $0$ in their multiplets.
Note that the top component of $f$-type vertex operator \cite{BW} 
provides this scalar $\overline{C}$ of helicity 0 while 
 the bottom  component of $f$-type vertex operator gives 
a graviton $e_2$ of maximum helicity 2.

Finally, the subamplitude we get, after 
taking account of all the numerical factors correctly, 
is given by 
\bea
(-i)^3 \langle r, s \rangle^4 \frac{[1,2][2,3][3,1]}{\langle 1,2
\rangle \langle 2,3\rangle \langle 3,1 \rangle}.
\label{result1}
\eea
Of course, after we assign the particles $r$ and $s$,
this can be further simplified. 

Similarly, 
for the subamplitude where
$\mu$-derivative  
of vertex operator at 
$1^{st}$ particle acts on the wavefunction of $3^{rd}$ particle, 
$\mu$-derivative  
of vertex operator at  
$3^{rd}$ particle acts on the wavefunction of $2^{nd}$ particle, 
and finally $\mu$-derivative  
of vertex operator at  
$2^{nd}$ particle acts on the wavefunction of $1^{st}$ 
particle(or equivalently by interchanging 
the roles of $2^{nd}$ particle 
and $3^{rd}$ particle:($2 \leftrightarrow 3$)), one gets,
by following above procedure,
\bea
(-i)^3 \langle r, s \rangle^4 \frac{[1,3][2,1][3,2]}{\langle 1,3
\rangle \langle 2,1\rangle \langle 3,2 \rangle}
\label{result2}
\eea
with a little rearrangement.

Next, let us consider the subamplitude where
the $\mu$-derivatives of vertex operator at 
both $1^{st}$ particle and $3^{rd}$ particle 
act on the wavefunction of $2^{nd}$ particle simultaneously.
Moreover, 
the $\mu$-derivative of vertex operator at 
$2^{nd}$ particle  
acts on the wavefunction of $3^{rd}$ particle.
In this case, since there exists only 
one spinor of negative helicity $\widetilde{\pi}_1^{\dot{a}}$ 
at $1^{st}$ particle, compared with previous cases, 
the cubic in $\pi_1^a$(the degree of $\la_1$ is three) 
of the vertex operator
at $1^{st}$ particle 
appears as a prefactor. Similarly, for the $2^{nd}$ particle,
only linear term in $\pi_2^a$(the degree of $\la_2$ is one) 
of the vertex operator
at $2^{nd}$ particle arises as a prefactor because there is a
cubic in $\widetilde{\pi}_2^{\dot{a}}$ after differentiating the 
exponentials.
For the $3^{rd}$ particle, the prefactor is the same as before:
quadratic in $\pi_3^a$(the degree of $\la_3$ is two).
The corresponding spinor indices in these prefactors
are contracted properly between 
these particles.  
Note that although the homogeneity in both $\la_1$ and $\la_2$ 
is not equal to zero in the 
amplitude(due to the different prefactors which are
not quadratic in $\pi_i^a$ and there is no appearance for 
$\mu_1$ derivative), 
compared with previous cases, 
the sum of degrees is zero.
With this information, one can write the subamplitude as
\bea
\int d^4 x d^8 \theta dz_1 dz_2 dz_3 
&&  \left(\frac{\pi_1^{a} \pi_{1b} \pi_{1c}}
{\pi_1^1 \pi_1^1 \pi_1^1} \right)
\de (\langle\la_1, \pi_1\rangle) 
\left(\la_1/\pi_1 \right)^2
\widetilde{\pi}_1^{\dot{a}} \frac{\pa}{\pa \mu_2^{\dot{a}}} 
 \frac{\pa}{\pa \mu_2^{\dot{c}}} 
\exp \left(
i (\pi_2/\la_2) [\mu_2,
  \widetilde{\pi}_2] 
\right)
 \nonu \\
&&  \left(\frac{\pi_{2a}}{\pi_2^1} \right) 
\de (\langle\la_2, \pi_2\rangle) 
\left(\la_2/\pi_2 \right)^2 
\widetilde{\pi}_2^{\dot{b}} \frac{\pa}{\pa \mu_3^{\dot{b}}} 
\exp \left(
i (\pi_3/\la_3) [\mu_3,
  \widetilde{\pi}_3] 
\right)
 \nonu \\
&&   \left(\frac{\pi_3^{b} \pi_3^{c}}{\pi_3^1 \pi_3^1}\right) 
\de (\langle\la_3, \pi_3\rangle) 
\left(\la_3/\pi_3 \right)^2 
\widetilde{\pi}_3^{\dot{c}}
\exp \left(
i (\pi_1/\la_1) [\mu_1,
  \widetilde{\pi}_1] 
\right)
 \nonu \\
&& 
\left\langle  
\prod_{k=1}^3 T(z_k) 
\right\rangle
 \prod_{j=1}^3 u \left((\pi_j/\la_j) \psi_j \right). 
\nonu
\eea
Then,
the factor $[1,2][2,3][3,2]$ for the spinors of
negative chirality arises in this amplitude by keeping track of 
both spinor index in $\widetilde{\pi}^{\dot{a}}$
and particle index in $\mu_j$ appropriately. 
Using the second relation of (\ref{relations}), 
the factor $z_{12} z_{31} z_{31}$ 
in the denominator of the correlator $\langle  
T(z_1) T(z_2) T(z_3) 
\rangle$ is removed. 
As a result, only $\frac{1}{z_{12} z_{23}^2}$ survives in this 
correlator.
We expect that 
there should be $1/(\pi_1^1)^2$ since in the prefactor there is an
extra factor $1/\pi_1^1$ and there is no $\mu_1$-derivative 
implying another extra factor $1/\pi_1^1$.  Similarly, 
$(\pi_2^1)^2$ will appear in the subamplitude because 
the effects from 
extra $\mu_2$-derivative and a single prefactor 
will add this factor. 
Finally, by simplifying this,
the subamplitude can be written as 
\bea
(-i)^3 \langle r, s \rangle^4 \frac{[1,2][2,3][3,2] \langle 
2, \zeta \rangle^2}{\langle 1,2
\rangle \langle 2,3\rangle \langle 3,2 \rangle \langle 1, \zeta
\rangle^2}
\label{result3}
\eea
where a spinor $\zeta^a=(0,1)$ is introduced \cite{BW}:
$(\pi_1^1)^2=\langle 1, \zeta
\rangle^2$ and $(\pi_2^1)^2=\langle 2, \zeta
\rangle^2$. 
One can count the homogeneity $2$ of $\la_1$ for $1^{st}$ particle
above amplitude as follows:
the integration measure $dz_1$ has 2 as in the discussion of
(\ref{deltacon}), the prefactor has 3, 
delta function has $-1$, the scale factor has 2, and the 
correlator has $-4$ from (\ref{tttcor}) of homogeneity in $\la_1$. 
This reflects on the final result (\ref{result3}) where 
there is a factor $(\pi_1^1)^2$ in the denominator by remembering 
(\ref{relations}).  
Similarly, one can count the homogeneity of 
$\la_2$ for the $2^{nd}$ particle
which is equal to $-2$. This is reminiscent of 
a factor $(\pi_2^1)^2$ in the numerator
of (\ref{result3}).  
Here the sum of degree of $\la_1$ and degree of $\la_2$ vanishes.   

One can analyze the other cases by permutations between 
three particles. In fact,
there are other five cases coming from the permutations between 
1, 2 and 3:$(132),(312),(213),(231)$, and $(321)$. 
Then the remaining scattering amplitude can be written as
\bea
(-i)^3 \langle r, s \rangle^4 \left(
\frac{[1,3][2,3][3,2] \langle 
3, \zeta \rangle^2}{\langle 1,3
\rangle \langle 2,3\rangle \langle 3,2 \rangle \langle 1, \zeta
\rangle^2} + \mbox{four other terms} \right)
\label{result4}
\eea
where the first term is obtained by interchanging the index 2 and 
index 3
from (\ref{result3}) with a rearrangement.  
Now it is ready to collect all the subamplitudes and sum over them.
Combining (\ref{result1}), (\ref{result2}), (\ref{result3})
and five other cases  characterized by 
(\ref{result4})(obtained by permutations), 
one writes
the full three particle scattering amplitude consisting of 
eight terms($2^3=8$) as a simplified and symmetric form 
\bea
(-i)^3 \langle r, s \rangle^4 \prod_{j=1}^{3} \sum_{k(\neq j)=1}^{3}
\frac{[j,k] \langle 
k, \zeta \rangle^2}{\langle j,k
\rangle  \langle j, \zeta
\rangle^2}.
\label{Ansthreeg}
\eea
This is exactly an expression for the MHV amplitude 
\cite{BW} for scattering
of three particles: a graviton of helicity 2 and two 
particles  labeled by $r$ and $s$ which are scalars  
$\overline{C}$ of helicity 0. 

So far, we have constructed the vertex operator at each 
particle and computed the amplitudes by executing the 
integrals. In doing this, we have to introduce a prefactor
for each particle which is a function of $\pi_i^a$
as well as both the correlator of stress-energy tensor and 
the wavefunction of plane wave.
The determination for this prefactor can be understood 
from the homogeneity of the sum of degrees for $\la_i$'s.   
Can we find $N$ particle scattering amplitude?
We will describe four particle case next.

$\bullet$ Four-point amplitude for four gravitons 

Now let us consider four graviton scattering amplitude.
One can compute four-point function 
from the results for three-point function(and their derivatives) and 
two-point function we have obtained(We will not present all 
the details here. Refer to \cite{Ragoucy} for the details).
It turns out to be 
\bea
\langle T(z_1) T(z_2) T(z_3) T(z_4) \rangle & = & 
c^2 \left( \frac{1}{z_{12}^4 z_{34}^4} + 
\frac{1}{z_{13}^4 z_{24}^4} +\frac{1}{z_{14}^4 z_{23}^4} \right) 
\nonu \\
&& + \frac{2c}{z_{23}^2 z_{34}^2 z_{24}^2} \left( 
\frac{1}{z_{12}^2} +
\frac{1}{z_{13}^2} +\frac{1}{z_{14}^2} -
\frac{1}{z_{12} z_{14}} -\frac{1}{z_{13} z_{14}} -
\frac{1}{z_{12} z_{13} }\right)
\nonu \\
& = & \frac{1}{z_{13}^4 z_{24}^4} \left[ 
c^2 \left( 1 + \frac{1}{x^4} +
\frac{1}{(1-x)^4}\right) + 2c \frac{(1-x+x^2)}{x^2(1-x)^2} \right]
\nonu \\
& \equiv &  \frac{1}{z_{13}^4 z_{24}^4} G
\left(\begin{array}{cc} 2 & 1 \\ 3 & 4 \end{array} \right)
(x)
\label{ttttcor}
\eea
where $z_{ij} \equiv z_i -z_j$, $x \equiv \frac{z_{12} z_{34}}
{z_{13} z_{24}}$ and $1-x \equiv \frac{z_{14} z_{23}}
{z_{13} z_{24}}$.
The quadratic terms in $c$ 
come from the sum of products of two-point functions
$\langle T(z_i) T(z_j) \rangle \langle T(z_k) T(z_l)\rangle$:
disconnected piece.
Note that due to the overall factor for linear term in $c$
\bea
\frac{1}{z_{13}^4 z_{24}^4} \frac{1}{x^2(1-x)^2} =
\frac{1}{z_{12}^2 z_{23}^2 z_{34}^2 z_{41}^2},
\nonu
\eea
the linear term of $c$ in four-point function is given by 
two terms: 
\bea
 \frac{2c}{z_{12} z_{23} z_{34} z_{41}}
 \left( \frac{1}{z_{12} z_{23} z_{34} z_{41}}+ 
\frac{1}{ z_{13}^2 z_{24}^2} \right)
\rightarrow 
 \frac{2c}{z_{12}^2 z_{23}^2 z_{34}^2 z_{41}^2}
\label{ttttcortrue}
\eea
where the first term behaves 
like the one given in three point function \footnote{One can write 
$z_{13} z_{24}$ in terms of other variables. That is, 
$z_{13} z_{24}= z_{12} z_{34}-z_{23} z_{41}$. Then the second term
of (\ref{ttttcortrue}) reduces to the first term of 
(\ref{ttttcortrue}) 
if we ignore double contractions $z_{i(i+1)}^2$ where $i=1, \cdots,
4$. }.
Even at the linear level for $c$, there 
exist two different contributions. We only consider 
the contribution from the first term above when we insert 
the correlator for stress-energy tensor into the scattering 
amplitude.
Recall that for current correlator, single-trace part is
proportional to $k$ while multi-trace part is proportional 
to higher powers of $k$.  
The crossing symmetry conditions(for example, 
\cite{BPZ,Ginsparg,BS}) 
that correspond to a change of
channels are characterized by
\bea
G 
\left(\begin{array}{cc} 2 & 1 \\ 3 & 4 \end{array} \right)
(x)  & = & (-1)^{h_1+h_2+h_3+h_4} x^{-2h_3} G 
\left(\begin{array}{cc} 2 & 4 \\ 3 & 1 \end{array} \right)
\left(\frac{1}{x}\right), \nonu \\
\qquad
G 
\left(\begin{array}{cc} 2 & 1 \\ 3 & 4 \end{array} \right)
(x) & = & (-1)^{h_1+h_2+h_3+h_4}  G 
\left(\begin{array}{cc} 4 & 1 \\  3 & 2 \end{array} \right)
(1-x)
\label{crossing}
\eea
where $h_i$ is a conformal dimension for $T(z_i)$.
In this case, $h_i=2$ where $i=1, \cdots, 4$. 
It can be easily checked that the conditions 
(\ref{crossing}) for crossing symmetry with explicit form 
(\ref{ttttcor})
are indeed 
satisfied.

Let us first consider the subamplitude where
the $\mu$-derivative $\frac{\pa}{\pa \mu_2}$ 
of vertex operator at  
$1^{st}$ particle acts on the wavefunction of $2^{nd}$ particle
and 
the $\mu$-derivative $\frac{\pa}{\pa \mu_1}$ 
of vertex operator at  
$2^{nd}$ particle acts on the wavefunction of $1^{st}$ particle.
The adjoining two spinor indices $\dot{a}$ appearing them(and
$\dot{b}$) are contracted each other.
Moreover, 
the $\mu$-derivative  
of vertex operator at  
$3^{rd}$ particle acts on the wavefunction of 
$4^{th}$ particle and vice versa.
As we have done for three-point amplitude, 
let us introduce a prefactor 
$\left(\frac{\pi_i^a \pi_{i}^{b}}{\pi_i^1 \pi_i^1}\right)$, which is 
obviously homogeneous of degree 0 in $\pi$,
in the vertex operator at 
the $i^{th}$ particle(Note that there exists 
a quadratic term $\widetilde{\pi}_i^{\dot{a}} 
\widetilde{\pi}_i^{\dot{b}}$  
for each $i^{th}$ particle after differentiating the 
exponentials).
As we have done before, the integrand of 
vertex operator contains 
$
f^{\dot{a}}_1 \frac{\pa}{\pa \mu^{\dot{a}}_2}
f^{\dot{b}}_2 \frac{\pa}{\pa \mu^{\dot{b}}_1}
f^{\dot{c}}_3 \frac{\pa}{\pa \mu^{\dot{c}}_4}
f^{\dot{d}}_4 \frac{\pa}{\pa \mu^{\dot{d}}_3}$
as well as both 
the correlator and prefactors.
Let us write the amplitude  explicitly by substituting 
the form of the wavefunctions (\ref{f}) as follows:
\bea
&& 
\left( \frac{\pi_{1c} \pi_{1d}}{\pi_1^1 \pi_1^1} \right)
\left( \frac{\pi_{2}^{a} \pi_{2}^{b}}{\pi_2^1 \pi_2^1} \right) 
\left( \frac{\pi_{3a} \pi_{3b}}{\pi_3^1 \pi_3^1} \right) 
\left( \frac{\pi_{4}^{c} \pi_{4}^{d}}{\pi_4^1 \pi_4^1} \right)  
\left\langle \prod_{k=1}^{4} T(z_k) \right\rangle
\left( \prod_{j=1}^4  
  \left(\la_j/\pi_j \right)^2
\de (\langle\la_j, \pi_j \rangle) 
u \left( \left(\pi_j/\la_j\right) \psi_j \right) \right)
\nonu \\
&& \widetilde{\pi}_1^{\dot{a}} \frac{\pa}{\pa \mu_2^{\dot{a}}} 
\exp \left(
i (\pi_2/\la_2) [\mu_2,
  \widetilde{\pi}_2] 
\right) 
\widetilde{\pi}_2^{\dot{b}} \frac{\pa}{\pa \mu_1^{\dot{b}}} 
\exp \left(
i (\pi_1/\la_1) [\mu_1,
  \widetilde{\pi}_1] 
\right)
\nonu \\
&&
\widetilde{\pi}_3^{\dot{c}} \frac{\pa}{\pa \mu_4^{\dot{c}}} 
\exp \left(
i (\pi_4/\la_4) [\mu_4,
  \widetilde{\pi}_4] 
\right)
\widetilde{\pi}_4^{\dot{d}} \frac{\pa}{\pa \mu_3^{\dot{d}}} 
\exp \left(
i (\pi_3/\la_3) [\mu_3,
  \widetilde{\pi}_3] 
\right).
\nonu
\eea
After differentiating the exponentials 
with respect to each $\mu$'s, 
the factor $[1,2][2,1][3,4][4,3]$ for the spinors of
negative chirality arises in this amplitude \footnote{
Here we ignored the integrals over $x_{a\dot{a}}, \theta^{Aa}$ and
$z_i$ where $i=1,2,3,4$ for simplicity.}. 
Using the second relation of (\ref{relations}), 
the factor $z_{23}^2 z_{41}^2$ 
in the denominator of the correlator $\langle  
\prod_{k=1}^{4} T(z_k)
\rangle$, characterized by (\ref{ttttcor}) or 
(\ref{ttttcortrue}), is removed
due to the presence of prefactors.
Then the remaining factor $\frac{1}{z_{12}^2 z_{34}^2}$,
that can be reexpressed as $\frac{1}{\langle 1,2
\rangle \langle 2,1\rangle \langle 3,4 \rangle \langle 4, 3 
\rangle}$ with some function of 
$\pi_i^1$'s, survives in this correlator.
It is easy to see that the homogeneity for each $\la_i$ is 
equal to zero: quadratic prefactors and 
all the $\mu_j$-derivatives appear once.   
As a result,  the factor $\langle1,2 \rangle \langle2,1 \rangle
\langle 3,4\rangle \langle 4,3 \rangle$ for the spinors of
positive chirality appears in the denominator of this 
subamplitude. 
Therefore, it turns out that the subamplitude can be written as 
\bea
(-i)^4 \langle r, s \rangle^4 
\frac{[1,2][2,1][3,4][4,3]}{\langle 1,2
\rangle \langle 2,1\rangle \langle 3,4 \rangle \langle 4, 3 
\rangle}.
\label{res1}
\eea
Also, there are two other cases where we simply replace 
the role of $2^{nd}$ particle with the one for $3^{rd}$ 
particle(can be interpreted as an interchanging of 
$2 \leftrightarrow 3$) and 
for second case, we replace $2^{nd}$ particle with $4^{th}$
particle($2 \leftrightarrow 4$). That is, the subamplitudes are
given by
\bea
(-i)^4 \langle r, s \rangle^4 
\frac{[1,3][2,4][3,1][4,2]}{\langle 1,3
\rangle \langle 2,4\rangle \langle 3,1 \rangle \langle 4, 2 
\rangle}, \qquad
(-i)^4 \langle r, s \rangle^4 
\frac{[1,4][2,3][3,2][4,1]}{\langle 1,4
\rangle \langle 2,3\rangle \langle 3,2 \rangle \langle 4, 1 
\rangle}
\label{res2}
\eea  
respectively.
When we make a permutation between the indices
1,2,3, and 4, there exists $4!=24$ possibilities. If we apply 
this permutation to (\ref{res1}), then there exist eight of 
(\ref{res1}), eight of first expression of (\ref{res2}), and 
eight of second expression of (\ref{res2}). 

Next, let us consider the 
particular amplitude
where the twistor derivative with respect to $\mu$ acts on 
$2^{nd}, 3^{rd}, 4^{th}$ and 
$1^{st}$ particles successively 
which is a generalization of (\ref{122331})
for four particle scattering and the amplitude can be written as,
by simply adding an extra $4^{th}$ particle,
\bea
&& 
\left( \frac{\pi_1^{a} \pi_{1d}}{\pi_1^1 \pi_1^1} \right)
\left( \frac{\pi_2^{b} \pi_{2a}}{\pi_2^1 \pi_2^1} \right) 
\left( \frac{\pi_3^{c} \pi_{3b}}{\pi_3^1 \pi_3^1} \right) 
\left( \frac{\pi_4^{d} \pi_{4c}}{\pi_4^1 \pi_4^1} \right)  
\left\langle \prod_{k=1}^{4} T(z_k) \right\rangle
\nonu \\
&& \prod_{j=1}^4  
 \left(\la_j/\pi_j \right)^2
\de (\langle\la_j, \pi_j \rangle) 
u \left( \left(\pi_j/\la_j\right) \psi_j \right)
\widetilde{\pi}_j^{\dot{a}_j} \frac{\pa}{\pa \mu_{j+1}^{\dot{a}_j}} 
\exp \left(
i  (\pi_{j+1} / \la_{j+1}) [\mu_{j+1},
  \widetilde{\pi}_{j+1}] 
\right).
\nonu
\eea
The homogeneity for each $\la_i$ is 
equal to zero: quadratic prefactors and 
all the $\mu_j$-derivatives appear once.  
In this case, after writing all the numerical factors correctly,
it is easy to see
the result can be written as
\bea
(-i)^4 \langle r, s \rangle^4 
\frac{[1,2][2,3][3,4][4,1]}{\langle 1,2
\rangle \langle 2,3\rangle \langle 3,4 \rangle \langle 4, 1 
\rangle}.
\label{res3}
\eea
Also there are five other cases by changing the role of 
each particle.
In other words, 
one can easily see that when we make a permutation for the 
expression (\ref{res3}), we get six different combinations 
including the case (\ref{res3}) with 
multiplicity four. 

So far, all the amplitudes we have obtained 
are homogeneous of degree 0 in 
each $\la_i$. 
That is, there is no $\zeta$ dependence in (\ref{res1}), 
(\ref{res2}), and
(\ref{res3}). 
Now let us consider 
the other cases where each degree for $\la_i$ is nonzero, 
in general, but the sum of degrees of $\la_i$'s 
is equal to zero.
Let us describe the subamplitude where
the $\mu$-derivatives of vertex operator at 
both $1^{st}$ particle and $3^{rd}$ particle 
acts on the wavefunction of $2^{nd}$ particle
and 
the $\mu$-derivatives of vertex operator at 
both $2^{nd}$ particle and $4^{th}$ particle 
acts on the wavefunction of $1^{st}$ particle.
One can determine the prefactors by requiring 
the sum of the number of $\pi_i$ and $\widetilde{\pi}_i$
should be equal to 4.
The subamplitude becomes
\bea
&& 
\left( \frac{\pi_{1d} }{\pi_1^1 } \right)
\left( \frac{\pi_2^{a} }{\pi_2^1 } \right) 
\left( \frac{\pi_3^{b}  \pi_3^c \pi_{3a}}{\pi_3^1 \pi_3^1 
\pi_3^1} \right) 
\left( \frac{\pi_4^{d} \pi_{4b} \pi_{4c}}{\pi_4^1 \pi_4^1 
\pi_4^1} \right)  
F(z_k, \pi_j, \la_j,\psi_j)
\nonu \\
&&
\widetilde{\pi}_1^{\dot{a}}  
\widetilde{\pi}_3^{\dot{c}} 
\frac{\pa}{\pa \mu_{2}^{\dot{a}}}
\frac{\pa}{\pa \mu_{2}^{\dot{c}}} 
\exp \left(
i  (\pi_{2} / \la_{2}) [\mu_{2},
  \widetilde{\pi}_{2}] 
\right)
\widetilde{\pi}_2^{\dot{b}}
\widetilde{\pi}_4^{\dot{d}} 
 \frac{\pa}{\pa \mu_{1}^{\dot{b}}} 
\frac{\pa}{\pa \mu_{1}^{\dot{d}}} 
\exp \left(
i  (\pi_{1} / \la_{1}) [\mu_{1},
  \widetilde{\pi}_{1}] 
\right)
\nonu
\eea
where we introduce a new notation which appears several times
later
\bea
F(z_k, \pi_j, \la_j,\psi_j) \equiv
\left\langle \prod_{k=1}^{4} T(z_k) \right\rangle
\prod_{j=1}^4  
\left(
 \la_j/\pi_j \right)^2
\de (\langle\la_j, \pi_j \rangle) 
u \left( \left(\pi_j/\la_j\right) \psi_j \right).
\label{largeF}
\eea
From above integrand, 
one can read off the factor $[1,2][2,1][3,2][4,1]$
by keeping track of the structure of spinor index in 
$\widetilde{\pi}^{\dot{a}}$ 
and 
particle index in $\mu_j$
and also 
the prefactors lead to the fact that  
the factor $z_{23} z_{34}^2 z_{41}$, in the 
denominator of the correlator 
for stress-energy tensor, is canceled out.
The fact that 
there are
one extra $\mu_1$-derivative, compared with previous case,  
and linear prefactor for
the $1^{st}$ particle 
allows us to have an extra $(\pi_1^1)^2$
in the subamplitude. 
This holds for the $2^{nd}$ particle.
On the other hand, for the $3^{rd}$ and $4^{th}$
particles, there are no $\mu_j$-derivative($j=3,4$) and cubic 
prefactors. This will lead to a factor $1/(\pi_3^1 \pi_4^1)^2$
in the amplitude. 
The final expression for this subamplitude is 
\bea
(-i)^4 \langle r, s \rangle^4 \frac{[1,2][2,1][3,2][4,1] \langle 
1, \zeta \rangle^2 \langle 2, \zeta \rangle^2}{\langle 1,2
\rangle \langle 2,1\rangle \langle 3,2 \rangle \langle 4, 1\rangle
\langle 3, \zeta
\rangle^2 \langle 4, \zeta
\rangle^2 }
\label{res4}
\eea
which reflects on the fact that this amplitude is homogeneous of 
degree $-2$ in 
$\la_1$, $-2$ in $\la_2$, 2 in $\la_3$ and 2 in $\la_4$. 
Therefore, 
the sum of degrees is equal to zero.
In this case also, 
the other cases can be obtained from a permutation on (\ref{res4}).
Actually, there are twelve different amplitudes including 
(\ref{res4}) with 
multiplicity 2.  

So far, 
we have only considered the cases where
any inner product $[i,j]$ in the subamplitude is a
contraction between two consecutive indices $i$
and $j(=i+1)$.
From now on, 
we will describe the subamplitudes 
for which there is a factor $[i,j]$ where $j=i+2$
as well as the factors $[i,i+1]$.

Let us consider the subamplitude where
the $\mu$-derivatives of vertex operator at 
both $3^{rd}$ particle and $4^{th}$ particle 
act on the wavefunction of $1^{st}$ particle, 
the $\mu$-derivative of vertex operator at 
$1^{st}$ particle acts on the wavefunction of 
$4^{th}$ particle, and 
the $\mu$-derivative of vertex operator at 
$2^{nd}$ particle acts on the wavefunction of 
$3^{rd}$ particle.
Since the correlator 
$\left\langle \prod_{k=1}^{4} T(z_k) 
\right\rangle$ does not contain $z_{13}$ at all,
as a prefactor, the term $
\left( \frac{\pi_{3}^{a}  \pi_{1a}}{\pi_3^1 
\pi_1^1} \right)^{-1}$
should be present in the amplitude, compared with
previous cases.
Still effectively,  
the  sum of the number of $\pi_i$ and $\widetilde{\pi}_i$
should be equal to 4.
The quadratic terms in $\pi_i^a$ for $i=1,4$
and cubic terms in $\pi_i^a$ for $i=2,3$ occur.
With this in mind, one can write the amplitude with 
(\ref{largeF})
as follows:
\bea
&& 
\left( \frac{\pi_{1}^{a} \pi_{1}^{b} }{\pi_1^1 \pi_1^1 } \right)
\left( \frac{\pi_{2}^c \pi_{2a}\pi_{2b} }{\pi_2^1
\pi_2^{1}\pi_2^{1} } \right) 
\left( \frac{\pi_3^{d}  \pi_3^e \pi_{3c}}{\pi_3^1 \pi_3^1 
\pi_3^1} \right) 
\left( \frac{\pi_{4d}  \pi_{4e}}{\pi_4^1 
\pi_4^1} \right)  
\left( \frac{\pi_{3}^{f}  \pi_{1f}}{\pi_3^1 
\pi_1^1} \right)^{-1}  
\widetilde{\pi}_1^{\dot{a}}  
\frac{\pa}{\pa \mu_{4}^{\dot{a}}} 
\exp \left(
i  (\pi_{4} / \la_{4}) [\mu_{4},
  \widetilde{\pi}_{4}] 
\right)
\nonu  \\
&&
\widetilde{\pi}_2^{\dot{b}} 
\frac{\pa}{\pa \mu_{3}^{\dot{b}}} 
\exp \left(
i  (\pi_{3} / \la_{3}) [\mu_{3},
  \widetilde{\pi}_{3}] 
\right) 
\widetilde{\pi}_3^{\dot{c}}
\widetilde{\pi}_4^{\dot{d}} 
 \frac{\pa}{\pa \mu_{1}^{\dot{c}}} 
\frac{\pa}{\pa \mu_{1}^{\dot{d}}} 
\exp \left(
i  (\pi_{1} / \la_{1}) [\mu_{1},
  \widetilde{\pi}_{1}] 
\right)
F(z_k, \pi_j, \la_j,\psi_j).
\nonu 
\eea
It is easy to see that 
the only factor $\frac{1}{z_{14} z_{23} z_{41}}$
in the correlator with the help of
(\ref{relations}) remains because the four prefactors 
cancel out $z_{12}^2 z_{23} z_{34}^2$.
For $3^{rd}$ particle, the overall factor has 
$1/(\pi_3^1)^2$ and there is no extra factor for this in the 
final expression.
For $4^{th}$ particle, the same thing happens.
However, for $1^{st}$ and $2^{nd}$ particles, 
the extra $\pi_1^1$ and $\pi_2^1$ dependence occurs.   
From this consideration, 
one can simplify the subamplitude as
\bea
(-i)^4 \langle r, s \rangle^4 \frac{[1,4][2,3][3,1][4,1] \langle 
1, \zeta \rangle^2}{\langle 1,4
\rangle \langle 2,3\rangle \langle 3,1 \rangle \langle 4, 1\rangle
\langle 2, \zeta
\rangle^2}.
\label{res5}
\eea
Also due to the permutations between four particles,
there are contributions from twenty three other terms.

Let us describe other subamplitude where
the $\mu$-derivatives of vertex operator at 
both $2^{nd}$ particle and $4^{th}$ particle 
acts on the wavefunction of $1^{st}$ particle, 
the $\mu$-derivative of vertex operator at 
$1^{st}$ particle acts on the wavefunction 
$3^{rd}$ particle, and 
the $\mu$-derivative of vertex operator at 
$3^{rd}$ particle acts on the wavefunction of $4^{th}$ particle.
In this case, the presence of 
factor $\widetilde{\pi}_1^{\dot{a}} 
\widetilde{\pi}_{3\dot{a}}$
needs to have an extra prefactor in the 
vertex operator. The $\pi_i^1$ dependence in the 
amplitude is the same as previous case:
same prefactors and same $\mu_j$-derivatives.
The subamplitude with (\ref{largeF}) 
is
\bea
&& 
\left( \frac{\pi_{1}^{a} \pi_{1e} }{ \pi_1^1 \pi_1^1} \right)
\left( \frac{\pi_{2}^b \pi_{2}^{c}\pi_{2a} }{\pi_2^1
\pi_2^{1}\pi_2^{1} } \right) 
\left( \frac{\pi_3^{d}   \pi_{3b} \pi_{3c}}{\pi_3^1
\pi_3^1  
\pi_3^1} \right) 
\left( \frac{\pi_{4}^{e}  \pi_{4d}}{\pi_4^1 
\pi_4^1} \right)  
\left( \frac{\pi_{3}^{f}  \pi_{1f}}{\pi_3^1 
\pi_1^1} \right)^{-1}  
\widetilde{\pi}_1^{\dot{a}}  
\frac{\pa}{\pa \mu_{3}^{\dot{a}}} 
\exp \left(
i  (\pi_{3} / \la_{3}) [\mu_{3},
  \widetilde{\pi}_{3}] 
\right)
\nonu \\
&&
\widetilde{\pi}_3^{\dot{c}} 
\frac{\pa}{\pa \mu_{4}^{\dot{c}}} 
\exp \left(
i  (\pi_{4} / \la_{4}) [\mu_{4},
  \widetilde{\pi}_{4}] 
\right) 
\widetilde{\pi}_2^{\dot{b}}
\widetilde{\pi}_4^{\dot{d}} 
 \frac{\pa}{\pa \mu_{1}^{\dot{b}}} 
\frac{\pa}{\pa \mu_{1}^{\dot{d}}} 
\exp \left(
i  (\pi_{1} / \la_{1}) [\mu_{1},
  \widetilde{\pi}_{1}] 
\right)
F(z_k, \pi_j, \la_j,\psi_j).
\nonu 
\eea
Therefore, the subamplitude is given by   
\bea
(-i)^4 \langle r, s \rangle^4 \frac{[1,3][2,1][3,4][4,1] \langle 
1, \zeta \rangle^2}{\langle 1,3
\rangle \langle 2,1\rangle \langle 3,4 \rangle \langle 4, 1\rangle
\langle 2, \zeta
\rangle^2}.
\label{res6}
\eea
By permutations between four particles, 
there are other contributions from 
twenty three terms.

Finally, let us describe
the subamplitude where 
the $\mu$-derivatives of vertex operator at 
$2^{nd}, 3^{rd}, 4^{th} $ particles act on 
the wavefunction of $1^{st}$ particle
and 
the $\mu$-derivative of vertex operator at 
$1^{st}$ particle acts on 
the wavefunction of $4^{th}$ particle.
Except $4^{th}$ particle, there is a dependence
on $\pi_i^1$($i=1,2,3$) in the amplitude.
The sum of degree of $\la_i$'s is zero.
Then the subamplitude is 
\bea
&& 
\left( \frac{\pi_{1}^{a}  }{ \pi_1^1 } \right)
\left( \frac{\pi_{2}^b \pi_{2}^{c}\pi_{2a} }{\pi_2^1
\pi_2^{1}\pi_2^{1} } \right) 
\left( \frac{\pi_3^{d} \pi_3^e \pi_{3b} \pi_{3c}}{\pi_3^1 \pi_3^1 
\pi_3^1 \pi_3^1} \right) 
\left( \frac{\pi_{4d}  \pi_{4e}}{\pi_4^1 
\pi_4^1} \right)  
\left( \frac{\pi_{3}^{f}  \pi_{1f}}{\pi_3^1 
\pi_1^1} \right)^{-1}  
F(z_k, \pi_j, \la_j,\psi_j)
\nonu \\
&&
\widetilde{\pi}_1^{\dot{a}}  
\frac{\pa}{\pa \mu_{4}^{\dot{a}}} 
\exp \left(
i  (\pi_{4} / \la_{4}) [\mu_{4},
  \widetilde{\pi}_{4}] 
\right)
\widetilde{\pi}_2^{\dot{b}} 
\widetilde{\pi}_3^{\dot{c}} 
\widetilde{\pi}_4^{\dot{d}} 
\frac{\pa}{\pa \mu_{1}^{\dot{b}}} 
\frac{\pa}{\pa \mu_{1}^{\dot{c}}} 
\frac{\pa}{\pa \mu_{1}^{\dot{d}}} 
\exp \left(
i  (\pi_{1} / \la_{1}) [\mu_{1},
  \widetilde{\pi}_{1}] 
\right). 
\nonu 
\eea
In this case, 
the subamplitude
can be written as
\bea
(-i)^4 \langle r, s \rangle^4 \frac{[1,4][2,1][3,1][4,1] \langle 
1, \zeta \rangle^4}{\langle 1,4
\rangle \langle 2,1\rangle \langle 3,1 \rangle \langle 4, 1\rangle
\langle 2, \zeta
\rangle^2 \langle 3, \zeta
\rangle^2 }.
\label{res7}
\eea
By permutations between 1,2,3, and 4, there exist also
eleven different subamplitudes.

Now it is ready to collect all the contributions from
subamplitudes we have considered. 
By combining all the contributions, 
(\ref{res1}), (\ref{res2}), (\ref{res3})(and corresponding 
five more terms), (\ref{res4})(and eleven more terms), 
(\ref{res5})(and twenty three more terms), 
(\ref{res6})(and twenty three more terms), 
and (\ref{res7})(and eleven more terms),
the full amplitude consisting of eighty one($3^4=81$) terms 
can be simplified as 
\bea
(-i)^4 \langle r, s \rangle^4 \prod_{j=1}^{4} 
\sum_{k(\neq j)=1}^{4}
\frac{[j,k] \langle 
k, \zeta \rangle^2}{\langle j,k
\rangle  \langle j, \zeta
\rangle^2}
\label{Ansfourg}
\eea
which is an expression for the MHV amplitude 
\cite{BW} for scattering
of four particles: two gravitons of helicity 2 and two 
particles  labeled by $r$ and $s$ which are scalars  
$\overline{C}$ of helicity 0. From the results (\ref{Ansthreeg}) and 
(\ref{Ansfourg}), one expects that the MHV amplitude for
scattering of $N$ gravitons can be extended by generalizing the 
summation and product index to $N$ with overall factor $(-i)^N$.

What happens for the system of graviton plus gluons?
We will describe them next.

\subsection{MHV tree amplitude for both
gluons and supergravitons  }

\indent

In this subsection, 
from the correlators between stress-energy tensor and 
current, we will describe 
MHV tree amplitudes for gravitons and gluons.
Let us first consider the three-point amplitude. 

$\bullet$ Three point amplitude for one graviton and two gluons

The OPE between stress-energy tensor 
and the current takes 
the standard form \cite{Ginsparg,BS}
\bea
T(z_1) J^{r_2}(z_2) & = & 
\frac{J^{r_2}(z_2)}{z_{12}^2}  
+\frac{   \pa J^{r_2}(z_2)}{z_{12}} + \cdots. 
\label{tjope}
\eea 
It is well-known that the 
current is a primary field of conformal dimension 
1.
From the OPE's (\ref{jjope}) and (\ref{tjope}),
the nonzero three-point function is given by 
two contributions from the contraction of $z_1$ and $z_2$ and 
the contraction of $z_1$
and $z_3$: 
\bea
\langle T(z_1) J^{r_2}(z_2) J^{r_3}(z_3) \rangle =k \de^{r_2 r_3} 
\left(
\frac{1}{z_{12}^2 z_{23}^2}-\frac{2}{z_{12} z_{23}^3}+
(2 \leftrightarrow 3)
\right)
\label{tjjcor}
\eea
where the second term which contains $\frac{1}{z_{23}^3}$ 
is due to the 
correlation function $\langle \pa J^{r_2}(z_2) J^{r_3}(z_3) 
\rangle$ and there is a trivial three point function 
coming from the fact that $\langle T(z_1) J^{r_2}(z_2) 
\rangle =0$:
\bea
\langle T(z_1) T(z_2) J^{r_3}(z_3) \rangle =0.
\nonu
\eea
Given the three-point function (\ref{tjjcor}),
how to construct the scattering amplitudes? Somehow the 
information for two gluons at $z_2$ and $z_3$ is encoded in 
(\ref{tjjcor}).

Let us compute the amplitude using the above three-point 
correlation function (\ref{tjjcor}).
First, let us describe the case where the $\mu$-derivative
at $1^{st}$ particle  acts on the wavefunction of $2^{nd}$ particle.
We expect to have $f^{\dot{a}}$ for the wavefunction of
supergraviton at $z_1$ and two $\phi$'s for the wavefunction of
two gluons at $z_2$ and $z_3$ in the integrand.
We introduce a factor $\left(\frac{\pi_i^a}{\pi_i^1}\right)$
which is homogeneous of degree 0 in $\pi_i$(does not give 
any scale in $\pi_i$)
and degree 1 in $\la_i$ by recalling (\ref{relations}).
This linear behavior is required by 
the vanishing of the sum of degrees of $\la_i$'s.
The vertex operator is given by $f_{1}^{\dot{a}} \frac{\pa}{\pa 
\mu_2^{\dot{a}}} \phi_2 \phi_3$ together with correlator 
$\langle  
T(z_1) J^{r_2}(z_2) J^{r_3}(z_3) 
\rangle$
and 
prefactors $\left(\frac{\pi_1^a}{\pi_1^1}\right) 
\left(\frac{\pi_{2a}}{\pi_2^1}\right)$.
This vertex operator scales as $t^{-4}$ for
$1^{st}$ particle and $t^{-2}$ for $2^{nd}$ and $3^{rd}$
particles, as we expected.  
The subamplitude by inserting the wavefunctions (\ref{phi}) 
and (\ref{f}) 
can be written as
\bea
\int d^4 x d^8 \theta dz_1 dz_2 dz_3 
&& 
\left(
\frac{\pi_1^a}{ \pi_1^1}  \right)
\de (\langle\la_1, \pi_1\rangle) 
\left(\la_1/ \pi_1 \right)^2
\widetilde{\pi}_1^{\dot{a}} \frac{\pa}{\pa \mu_2^{\dot{a}}} 
\exp \left(
i (\pi_2/ \la_2) [\mu_2,
  \widetilde{\pi}_2] 
\right)
 \nonu \\
&&  
\left(\frac{\pi_{2a}}{\pi_2^1} \right)  
\de (\langle\la_2, \pi_2\rangle) 
\left(\la_2 / \pi_2 \right)
\exp \left(
i (\pi_3 / \la_3) [\mu_3,
  \widetilde{\pi}_3] 
\right)
 \nonu \\
&&  \de (\langle\la_3, \pi_3\rangle) 
\left(\la_3/ \pi_3 \right) 
\exp \left(
i (\pi_1 / \la_1) [\mu_1,
  \widetilde{\pi}_1] 
\right)
 \nonu \\
&& 
\langle  
T(z_1) J^{r_2}(z_2) J^{r_3}(z_3) 
\rangle
 \prod_{j=1}^3 u \left((\pi_j/\la_j) \psi_j \right). 
\nonu
\eea
In the correlator (\ref{tjjcor}), there are two contributions, but
the correct term is related to a term $\frac{1}{z_{12}^2 z_{23}^2}$ 
since the amplitude for gluons($2^{nd}$ particle and $3^{rd}$
particle) 
should behave as $\frac{1}{z_{23}^2}$. 
The prefactor 
$\left( \frac{\pi_i^a}{ \pi_i^1}  \right)$
is homogeneous of degree 0 in $\pi$ and of degree 1 in $\la$.
After differentiating the exponential 
with respect to $\mu$, 
the factor $[1,2]$ for the spinors of
negative chirality arises in this amplitude. 
The dependence on $\pi_3^1$ disappears 
because the factor $1/(\pi_3^1)^2$ is canceled 
out $(\pi_3^1)^2$ from the correlator. 
Using the second relation of (\ref{relations}), 
the factor $z_{12}$ 
in the denominator of the correlator $\langle  
T(z_1) J^{r_2}(z_2) J^{r_3}(z_3)  
\rangle$, characterized by (\ref{tjjcor}), 
is removed from the prefactor. 
As a result,  the factor 
$\langle1,2 \rangle \langle2,3 \rangle^2$ for the spinors of
positive chirality appears in the denominator of this 
amplitude. 
This amplitude can be simplified as
\bea
(-i) \frac{1}{\langle 2,3 \rangle^2}
\langle r, s\rangle^4
\frac{[1,2]\langle 2, \zeta \rangle^2 }
{\langle 1,2 \rangle \langle 1,
\zeta \rangle^2}. 
\label{result-one}
\eea
Definitely, 
the sum of degrees of $\la_1$ and $\la_2$ is equal to zero.
The factor $1/\langle 2,3\rangle^2$ implies the 
MHV tree amplitude for two gluons(omitting the group theory
factor).

Moreover, one can compute 
the case where 
the $\mu$-derivative
at $1^{st}$ particle  acts on the 
wavefunction of 
$3^{rd}$ particle  instead of $2^{nd}$ particle.
The corresponding correlator for this case is
given by  $\frac{1}{z_{13}^2 z_{32}^2}$ 
from (\ref{tjjcor}). 
Therefore, the full amplitude coming from 
these two contributions becomes, by summing over
$2^{nd}$ and $3^{rd}$ particles,
\bea
(-i) \frac{1}{\langle 2,3 \rangle^2}
\langle r, s \rangle^4  \sum_{k=2}^{3}
\frac{[1,k] \langle 
k, \zeta \rangle^2}{\langle 1,k
\rangle  \langle 1, \zeta
\rangle^2}
\label{Ansthreegg}
\eea
which is an expression for 
the MHV tree amplitude \cite{BW} for scattering 
of three particles: a graviton of helicity 2 characterized by 
the $1^{st}$ particle and 
two particles which are gauge bosons(labeled by $r$ and $s$) 
of helicity $-1$. 

So far, we have constructed 
the vertex operator at each particle and computed 
the amplitudes by calculating the integrals.
The behavior of prefactor is linear and is determined by
the requirement that the sum of degrees of $\la_i$'s vanishes.
Now we continue to the higher correlation function next. 

$\bullet$ Four point amplitude for one graviton and three gluons

By considering the contractions between $z_1$ and $z_2,z_3,z_4$,
the four-point function
can be obtained and is given by  
\bea
  \langle T(z_1) J^{r_2}(z_2) J^{r_3}(z_3) 
J^{r_4}(z_4) \rangle 
   = 
\frac{f^{r_2 r_3 r_4}}{z_{12} z_{23} z_{34} z_{42}}
\left( \frac{1}{z_{12}} -\frac{1}{ z_{23}} +
\frac{1}{ z_{42}} \right) 
+ (2 \leftrightarrow 3) +(2
\leftrightarrow 4).
\label{tjjjcor}
\eea
Note that 
the overall factor of the first term has
a dependence on $\frac{1}{z_{23} z_{34} z_{42}}$ which can be seen
from three-point correlator of the current (\ref{jjjcor}) and only
this 
first term is relevant to our scattering amplitude. 

The scattering amplitude where the twistor derivative with 
respect to $\mu$ at $1^{st}$ particle acts on the wavefunction of
$2^{nd}$ particle which is simply an extension of 
(\ref{result-one}) for three gluons. 
The vertex operator is given by $f_{1}^{\dot{a}} \frac{\pa}{\pa 
\mu_2^{\dot{a}}} \phi_2 \phi_3 \phi_4$
as well as the correlator (\ref{tjjjcor}) and prefactors.
Then the subamplitude, by adding the $4^{th}$ particle, 
can be  written as
\bea
&& 
\left(
\frac{\pi_1^a}{ \pi_1^1}  \right)
\left(\frac{\pi_{2a}}{\pi_2^1} \right)  
\left(\la_1/ \pi_1 \right)^2
\widetilde{\pi}_1^{\dot{a}} \frac{\pa}{\pa \mu_2^{\dot{a}}} 
\exp \left(
i (\pi_2/ \la_2) [\mu_2,
  \widetilde{\pi}_2] 
\right)
\left(\la_2 / \pi_2 \right)
\exp \left(
i (\pi_3 / \la_3) [\mu_3,
  \widetilde{\pi}_3] 
\right)
 \nonu \\
&&  
\left(\la_3/ \pi_3 \right) 
\exp \left(
i (\pi_4 / \la_4) [\mu_4,
  \widetilde{\pi}_4] 
\right)  
\left(\la_4/ \pi_4 \right) 
\exp \left(
i (\pi_1 / \la_1) [\mu_1,
  \widetilde{\pi}_1] 
\right)
 \nonu \\
&& 
\langle  
T(z_1) \prod_{k=2}^{4} J^{r_k}(z_k)
\rangle
 \prod_{j=1}^4 
\de (\langle\la_j, \pi_j \rangle) 
u \left((\pi_j/\la_j) \psi_j \right)
\nonu
\eea
where we did not write down the integrals explicitly for simplicity.
In the correlator (\ref{tjjjcor}), the only first term is
relevant to our discussion because 
the amplitude for three gluons should behave like
$\frac{1}{z_{23} z_{34} z_{42}}$.
The factor $[1,2]$ for the spinors of
negative chirality arises in this amplitude by 
differentiating the exponential with respect to $\mu_2$. 
Using (\ref{relations}), 
the factor $z_{12}$ 
in the denominator of the correlator $\langle  
T(z_1) J^{r_2}(z_2) J^{r_3}(z_3)   J^{r_4}(z_4)
\rangle$, characterized by (\ref{tjjjcor}), is canceled
by the prefactors. 
As a result,  the factor 
$\langle1,2 \rangle \langle2,3 \rangle  \langle 3,4 \rangle
 \langle 4,2 \rangle $ 
for the spinors of
positive chirality appears in the denominator of this 
amplitude. 
Then the amplitude can be written as
\bea
(-i) \frac{1}{\langle 2,3 \rangle\langle 3,4 \rangle
\langle 4,2 \rangle }
\langle r, s\rangle^4
\frac{[1,2]\langle 2, \zeta \rangle^2 }
{\langle 1,2 \rangle \langle 1,
\zeta \rangle^2}. 
\nonu
\eea

One can proceed the other two cases 
where the role of $2^{nd}$ particle is replaced by
$3^{rd}$ and $4^{th}$ particles
and summing over these contributions,
one arrives at the full amplitude given by
\bea
(-i) 
\frac{1}{\langle 2,3 \rangle\langle 3,4 \rangle
\langle 4,2 \rangle }
\langle r, s \rangle^4  \sum_{k=2}^{4}
\frac{[1,k] \langle 
k, \zeta \rangle^2}{\langle 1,k
\rangle  \langle 1, \zeta
\rangle^2}.
\label{restjjj}
\eea
This is
the MHV tree amplitude \cite{BW} for scattering 
of four particles: a graviton of helicity 2, a gauge boson
of helicity 1  and 
two particles which are gauge bosons
of helicity $-1$. 

One can easily see the generalization of this scattering 
amplitude to $N$ external particles consisting of 
one graviton and $(N-1)$ gluons.
The result is obtained 
by summing over $k$ to $N$ in (\ref{restjjj})
and the prefactor has the general expression for $(N-1)$ 
gluons:$\prod_{i=2}^{N} \frac{1}{\langle i, i+1\rangle}$.

$\bullet$ Four point amplitude for two gravitons and two gluons

Let us consider the cases where there are two gravitons.
There exists a four-point function 
\bea
\langle T(z_1) T(z_2) T(z_3) J^{r_4}(z_4) \rangle =0
\nonu
\eea
which can be obtained from the vanishing of correlation functions
$\langle T(z_1) T(z_2) J^{r_3}(z_3) \rangle =0$ and 
$\langle T(z_1) J^{r_2}(z_2) \rangle =0$
and the following four-point function has 
rather  complicated expression 
\bea
\langle T(z_1) T(z_2) J^{r_3}(z_3) J^{r_4}(z_4) \rangle 
& = & \frac{c k \de^{r_3 r_4} /2}{z_{12}^4 z_{34}^2} \nonu \\
&+& 
 k \de^{r_3 r_4} 
\left( \frac{2}{z_{12}^2} + \frac{1}{z_{13}^2} +
\frac{1}{z_{14}^2} \right)
 \left(
\frac{1}{z_{23}^2 z_{34}^2}-\frac{2}{z_{23} z_{34}^3}+
(3 \leftrightarrow 4) \right) \nonu \\
& + & \mbox{other singular terms} 
\label{ttjjcor}
\eea
where the first term comes from the 
disconnected piece $\langle T(z_1) T(z_2) \rangle \langle
J^{r_3}(z_3) J^{r_4}(z_4) \rangle$ \footnote{
It is easy to see that the other singular terms 
are given by the following correlation functions
$
\frac{1}{z_{12}} 
\langle \pa T(z_2) J^{r_3}(z_3) J^{r_4}(z_4) \rangle
+ \frac{1}{z_{13}} 
\langle  T(z_2) \pa J^{r_3}(z_3) J^{r_4}(z_4) \rangle
+ \frac{1}{z_{14}} 
\langle  T(z_2) J^{r_3}(z_3) \pa J^{r_4}(z_4) \rangle$.}.
The term of $\frac{1}{z_{34}^2}$ in the second term
provides the correlator for two currents and is 
relevant to our discussion. 

Let us consider
the subamplitude where 
the $\mu$-derivative of vertex operator at 
$1^{st}$ particle  acts on 
the wavefunction of $2^{nd}$ particle
and
the $\mu$-derivative of vertex operator at 
$2^{nd}$ particle  acts on 
the wavefunction of $1^{st}$ particle.
In this case, the prefactors coming from 
the two gravitons are the same as the one we have 
discussed in the previous subsection around (\ref{res1}). 
There is no $\pi_i^1$ dependence in the amplitude.
We expect to have 
two $\phi$'s from two gluons. 
The integrand we are interested in is 
\bea
&& 
\left( \frac{\pi_1^a \pi_{1b} }{ \pi_1^1 \pi_1^1} \right)
\left( \frac{\pi_2^{b} \pi_{2a}}{\pi_2^1 \pi_2^1} \right)
H(z_k, \pi_j, \la_j,\psi_j)
\left(\la_1 / \pi_1 \right)^2
\widetilde{\pi}_1^{\dot{a}} 
\frac{\pa}{\pa \mu_2^{\dot{a}}} 
\exp \left(
i (\pi_2/ \la_2) [\mu_2,
  \widetilde{\pi}_2] 
\right)
\nonu \\
&&
\left(\la_2/ \pi_2 \right)^2 
\widetilde{\pi}_2^{\dot{b}}
\frac{\pa}{\pa \mu_1^{\dot{b}}} 
\exp \left(
i (\pi_1/ \la_1) [\mu_1,
  \widetilde{\pi}_1] 
\right)
\left(\la_3/ \pi_3 \right)
\exp \left(
i (\pi_3/ \la_3) [\mu_3,
  \widetilde{\pi}_3] 
\right)
\nonu \\
&&
\left(\la_4/ \pi_4 \right)
\exp \left(
i (\pi_4/ \la_4) [\mu_4,
  \widetilde{\pi}_4] 
\right)
\nonu
\eea
where we introduce a new notation
which will appear several times
\bea
H(z_k, \pi_j, \la_j,\psi_j)
\equiv
\langle  
T(z_1) T(z_2) J^{r_3}(z_3) J^{r_4}(z_4)
\rangle
  \prod_{j=1}^4 
\de (\langle\la_j, \pi_j \rangle) 
u \left((\pi_j/\la_j) \psi_j \right).
\label{largeH}
\eea
The relevant term in the correlator (\ref{ttjjcor})
is $\frac{1}{z_{12}^4  z_{34}^2}$ because
in this case the amplitude for gluons($3^{rd}$ particle and $4^{th}$
particle) 
behaves correctly.
After simplifying the above(the prefactors 
kill $z_{12}^2$) and extracting the first term for linear $k$ 
in the 
correlator (\ref{ttjjcor}), 
the subamplitude can be written 
\bea
(-i)^2 \frac{1}{\langle 3,4 \rangle^2 }
\langle r, s\rangle^4
 \frac{[1,2][2,1]}
{\langle 1,2 \rangle \langle 2,1 \rangle }.  
\label{resultone}
\eea
The factor $1/\langle 3,4\rangle^2$ implies the 
MHV tree amplitude for two gluons.

Let us consider
the subamplitude where 
the $\mu$-derivative of vertex operator at 
$1^{st}$ particle  acts on 
the wavefunction of $2^{nd}$ particle
and
the $\mu$-derivative of vertex operator at 
$2^{nd}$ particle  acts on 
the wavefunction of $3^{rd}$ particle.
The prefactor in $2^{nd}$ particle 
is quadratic in $\pi_2^a$ because the spinor indices
are contracted with the one in $1^{st}$ particle and the one in 
$3^{rd}$ particle.
Then with appropriate prefactors which will
remove some factors in the correlator (\ref{ttjjcor})
where the relevant term is 
$\frac{1}{z_{12}^2 z_{23}^2 z_{34}^2}$, 
the subamplitude 
with (\ref{largeH}) is 
\bea
&& 
\left( \frac{\pi_1^a }{ \pi_1^1} \right)
\left( \frac{\pi_2^{b} \pi_{2a}}{\pi_2^1 \pi_2^1} \right)
\left( \frac{\pi_{3b} }{ \pi_3^1} \right) 
H(z_k, \pi_j, \la_j,\psi_j)
\left(\la_1 / \pi_1 \right)^2
\widetilde{\pi}_1^{\dot{a}} 
\frac{\pa}{\pa \mu_2^{\dot{a}}} 
\exp \left(
i (\pi_2/ \la_2) [\mu_2,
  \widetilde{\pi}_2] 
\right)
\nonu \\
&&
\left(\la_2/ \pi_2 \right)^2 
\widetilde{\pi}_2^{\dot{b}}
\frac{\pa}{\pa \mu_3^{\dot{b}}} 
\exp \left(
i (\pi_3/ \la_3) [\mu_3,
  \widetilde{\pi}_3] 
\right) 
\left(\la_3/ \pi_3 \right)
\exp \left(
i (\pi_4/ \la_4) [\mu_4,
  \widetilde{\pi}_4] 
\right)
\nonu \\
&&
\left(\la_4/ \pi_4 \right)
\exp \left(
i (\pi_1/ \la_1) [\mu_1,
  \widetilde{\pi}_1] 
\right).
\nonu
\eea
It can be easily checked that 
the degree of $\la_1$ is 2 while the degree of $\la_3$
is $-2$ but the sum of degrees is equal to zero.
There is also other contribution by 
taking 
the $\mu$-derivative of vertex operator at 
$2^{nd}$ particle  which acts on 
the wavefunction of $4^{th}$ particle
instead of $3^{rd}$ particle($3 \leftrightarrow 4$).
In this case, we have to use the corresponding 
relevant term in (\ref{ttjjcor}).
Then the total subamplitude coming from two contributions
can be written as
\bea
(-i)^2 \frac{1}{\langle 3,4 \rangle^2 }
\langle r, s\rangle^4
 \left( \frac{[1,2][2,3]\langle 3, \zeta \rangle^2}
{\langle 1,2 \rangle \langle 2,3 \rangle 
\langle 1, \zeta \rangle^2 }  + (3 \leftrightarrow 4) \right).
\label{resulttwo}
\eea
Obviously, the sum of degrees for $\la_1$ and $\la_3$
is zero \footnote{
For
the subamplitude where 
the $\mu$-derivative of vertex operator at 
$1^{st}$ particle  acts on 
the wavefunction of $3^{rd}$ particle
and
the $\mu$-derivative of vertex operator at 
$2^{nd}$ particle  acts on 
the wavefunction of $1^{st}$ particle,
the prefactor of $1^{st}$ particle is
quadratic in $\pi_1^a$. 
The subamplitude is given by similarly.
By taking account of other case which 
changes the role of $3^{rd}$ particle($3 \leftrightarrow 4$),
one gets 
$
(-i)^2 \frac{1}{\langle 3,4 \rangle^2 }
\langle r, s\rangle^4
 \left( \frac{[1,3][2,1]\langle 3, \zeta \rangle^2}
{\langle 1,3 \rangle \langle 2,1 \rangle 
\langle 2, \zeta \rangle^2 }  + (3 \leftrightarrow 4) \right)$.
Actually this can be obtained from the result 
(\ref{resulttwo}) by exchanging the role of 
$1^{st}$ particle and $2^{nd}$ 
particle($1 \leftrightarrow 2$). This is the reason why the 
coefficient of $\frac{1}{z_{12}^2 z_{23}^2 z_{34}^2}$ in 
(\ref{ttjjcor}) is twice as
large as those of other terms.}.

Let us consider
the subamplitude where 
the $\mu$-derivatives of vertex operator at 
$1^{st}$ particle  and $2^{nd}$ particle acts on 
the wavefunction of $3^{rd}$ particle simultaneously.
The prefactor for $3^{rd}$ particle is quadratic 
and the prefactors for other particle are linear.
The result is 
\bea
&& 
\left( \frac{\pi_1^a}{\pi_1^1} \right)
\left( \frac{\pi_2^b}{\pi_2^1} \right)
\left( \frac{\pi_{3a} \pi_{3b}}{\pi_3^1 \pi_3^1} \right) 
H(z_k, \pi_j, \la_j,\psi_j)
\left(\la_1 / \pi_1 \right)^2
\left(\la_2/ \pi_2 \right)^2
\widetilde{\pi}_1^{\dot{a}} 
\widetilde{\pi}_2^{\dot{b}}
\nonu \\
&&
\frac{\pa}{\pa \mu_3^{\dot{a}}} 
\frac{\pa}{\pa \mu_3^{\dot{b}}} 
\exp \left(
i (\pi_3/ \la_3) [\mu_3,
  \widetilde{\pi}_3] 
\right)
\exp \left(
i (\pi_4/ \la_4) [\mu_4,
  \widetilde{\pi}_4] 
\right)
 \nonu \\
&&  
\left(\la_3/ \pi_3 \right) 
\exp \left(
i (\pi_1/ \la_1) [\mu_1,
  \widetilde{\pi}_1] 
\right)
\left(\la_4/ \pi_4 \right)
\exp \left(
i (\pi_2/ \la_2) [\mu_2,
  \widetilde{\pi}_2] 
\right).
\nonu
\eea
For $3^{rd}$ particle, 
the prefactor is quadratic and there are two 
$\mu_3$-derivatives. Then one can easily check that
the degree of $\la_3$ is equal to 4.
On the other hand, for $4^{th}$ particle,
the factor $1/(\pi_4^1)^2$ coming from both the scale above and 
delta function is canceled out $(\pi_4^1)^2$ in 
$1/z_{34}^2$.
For $1^{st}$ and $2^{nd}$ particles, the scales 
$1/(\pi_1^1)^2$ and $1/(\pi_2^1)^2$ remain in the 
amplitude. Finally, the sum of degrees for $\la_i$'s 
is equal to zero.
Note that the relevant term in the correlator (\ref{ttjjcor}) 
is $\frac{1}{z_{13}^2 z_{23}^2 z_{34}^2}$.
In this case, the subamplitude, by adding other 
contribution(by changing the role of $3^{rd}$ particle and
$4^{th}$ particle), can be summarized
\bea
(-i)^2 \frac{1}{\langle 3,4 \rangle^2 }
\langle r, s\rangle^4
\left(
 \frac{[1,3][2,3]\langle 3, \zeta \rangle^4}
{\langle 1,3 \rangle \langle 2,3 \rangle 
\langle 1, \zeta \rangle^2 \langle 2, \zeta \rangle^2 }  +
 (3 \leftrightarrow 4)
\right).
\label{resultfour}
\eea

Finally, let us describe
the subamplitude where 
the $\mu$-derivative of vertex operator at 
$1^{st}$ particle acts on 
the wavefunction of $3^{rd}$ particle
and 
the $\mu$-derivative of vertex operator at 
$2^{nd}$ particle acts on 
the wavefunction of $4^{th}$ particle.
All the prefactors in this case are linear in $\pi_i^a$ for
each $i^{th}$ particle.
The subamplitude with (\ref{largeH}) is given by 
\bea
&& 
\left( \frac{\pi_1^a}{\pi_1^1} \right)
\left( \frac{\pi_2^b}{\pi_2^1} \right)
\left( \frac{\pi_{3a}}{\pi_3^1} \right) 
\left( \frac{\pi_{4b}}{\pi_4^1} \right) 
H(z_k, \pi_j, \la_j,\psi_j)
\left(\la_1 / \pi_1 \right)^2
\widetilde{\pi}_1^{\dot{a}} \frac{\pa}{\pa \mu_3^{\dot{a}}} 
\exp \left(
i (\pi_3/ \la_3) [\mu_3,
  \widetilde{\pi}_3] 
\right)
 \nonu \\
&&
\left(\la_2/ \pi_2 \right)^2
\widetilde{\pi}_2^{\dot{b}} \frac{\pa}{\pa \mu_4^{\dot{b}}} 
\exp \left(
i (\pi_4/ \la_4) [\mu_4,
  \widetilde{\pi}_4] 
\right)
\left(\la_3/ \pi_3 \right) 
\exp \left(
i (\pi_1/ \la_1) [\mu_1,
  \widetilde{\pi}_1] 
\right)
 \nonu \\
&&   
\left(\la_4/ \pi_4 \right)
\exp \left(
i (\pi_2/ \la_2) [\mu_2,
  \widetilde{\pi}_2] 
\right).
\nonu
\eea
As we have done before, 
in this case, 
the dependence on $\pi_i^1$ where $i=1,2,3,4$ in this amplitude
occurs.
For $1^{st}$ and $2^{nd}$ particles, 
the scales $1/(\pi_1^1)^2$ and $1/(\pi_2^1)^2$
remain in the amplitude. 
On the other hand, for $3^{rd}$ and $4^{th}$ particles, 
the factors $(\pi_3^1)^2$ and $(\pi_4^1)^2$ appear.
The sum of degrees of $\la_i$'s is zero.
The relevant term in the correlator (\ref{ttjjcor})
is $\frac{1}{z_{13}^2 z_{24}^2 z_{34}^2}$ because
in this case the amplitude for gluons($3^{rd}$ particle and $4^{th}$
particle) 
behaves correctly.
After we take into account of other case where 
the role of $3^{rd}$ particle and $4^{th}$
particle is interchanged($3  \leftrightarrow 4$), 
the final expression is
given by
\bea
(-i)^2 \frac{1}{\langle 3,4 \rangle^2 }
\langle r, s\rangle^4
\left( 
\frac{[1,3][2,4]\langle 3, \zeta \rangle^2
\langle 4, \zeta \rangle^2}
{\langle 1,3 \rangle \langle 2,4 \rangle 
\langle 1, \zeta \rangle^2 \langle 2, \zeta \rangle^2 }  
+ (3 \leftrightarrow 4) \right).
\label{resultfive}
\eea

From the contributions (\ref{resultone}),
(\ref{resulttwo})(and two other terms), 
(\ref{resultfour}) and (\ref{resultfive}),
the full amplitude($3^2=9$) by rearrangement is given by 
\bea
(-i)^4 
\frac{1}{\langle 3,4 \rangle^2 }
\langle r, s \rangle^4 \prod_{j=1}^{2} \sum_{k(\neq j)=1}^{4}
\frac{[j,k] \langle 
k, \zeta \rangle^2}{\langle j,k
\rangle  \langle j, \zeta
\rangle^2}.
\label{final}
\eea
which is an expression for the MHV tree amplitude 
\cite{BW} for scattering
of four particles: two gravitons of helicity 2 and two 
gluons  labeled by $r$ and $s$ of helicity $-1$. 
The generalization of this scattering 
amplitude to $N$ external particles consisting of 
two gravitons and $(N-2)$ gluons can be obtained 
by summing over $k$ to $N$ in (\ref{final})
and the prefactor has the general expression for $(N-2)$ 
gluons.

\section{Discussions }
\setcounter{equation}{0}

\indent

We have obtained
three-point and four-point amplitudes for gravitons
(\ref{Ansthreeg}) and (\ref{Ansfourg}), 
three-point and four-point amplitudes for gravitons and 
gluons (\ref{Ansthreegg}), (\ref{restjjj}) and (\ref{final}).
These results agree with the description \cite{BW} from 
open string version. 
Can we generalize for arbitrary $N$-point($N \geq 5$) amplitudes
as we mentioned before? It is rather obvious that from the 
open string version of twistor-string theory, the MHV
tree scattering amplitudes is generalized to arbitrary $N$-point
amplitudes, starting from the scattering amplitude between 
any two different particles.

Now we will describe  the five-point 
amplitudes very briefly
in the context of topological B-model. 
Let us first consider the amplitude for five gravitons and 
we will add the gluons later. The latter is easier 
to analyze than the former. 

$\bullet$ Five-point amplitude for five gravitons 

Now let us consider the five-point function.
One can find this from the OPE's between stress-energy tensor
\footnote{
For example,  
by reading off the contractions between $z_1$ and $z_k$ where 
$k=1, \cdots, 4$, the quadratic terms of
$c$ in this correlator 
$\langle \prod_{k=1}^{5} T(z_k) \rangle$
are given by 
$
\frac{c^2}{2}  \frac{1}{z_{12}^4 z_{34}^2 
z_{45}^2 z_{53}^2} \left[ 1 +x^2 y^2 + 
\frac{x^2(x-y)^2}{(1-x)^4} 
+\frac{y^2 (x-y)^2 }{(1-y)^4} \right]
$
where we introduced the two variables
$
x \equiv \frac{z_{12} z_{34}}
{z_{13} z_{24}}, 
y \equiv \frac{z_{12} z_{35}}
{z_{13} z_{25}}$.}. 
Using the relation \footnote{
It is easy to check that $
\frac{1}{z_{12}^4 z_{34}^2 
z_{45}^2 z_{53}^2} = \frac{1}{z_{12}^2 z_{23}^2 z_{34}^2 z_{45}^2 
z_{51}^2} \left( \frac{1-y}{y}\right)^2$.},
the five-point function 
\footnote{The linear terms in $c$ of 
$\langle \prod_{k=1}^{5} T(z_k) \rangle$
can be summarized by
$
\frac{2c}{z_{12}^4 z_{34}^2 
z_{45}^2 z_{53}^2} \frac{1}{(1-x)^2(1-y)^2} F(x,y) 
$ where we introduce 
$F(x,y) \equiv  2x^4 
(1-y+y^2) +2y^2 (1-y+y^2) -x^3 (2+y+y^2+2y^3) 
-xy 
(2+y+y^2+2y^3) +x^2 (2-y+6y^2-y^3+2y^4)$.}
can be written as
\bea
 && 
\frac{2c}{z_{12}^2 z_{23}^2 
z_{34}^2 z_{45}^2 z_{51}^2} + \mbox{other singular terms}. 
\nonu
\eea
For the case of five gravitons, there exist $4^5 =1024$ terms
for the scattering amplitudes.
When we take the contraction of spinor indices 
between $\widetilde{\pi}$ and  
$\frac{\pa}{\pa \mu}$, there are following different 
inner products $[1,3],[2,4],[3,5],[4,1],[5,2]$ where the 
corresponding factors for the spinors of positive chirality 
$\langle 1,3 \rangle, \langle 2,4\rangle, \langle 3,5\rangle, 
\langle 4,1 \rangle,  \langle 5,2 \rangle$ are not present in 
the correlator $\prod_{k=1}^{5} \langle T(z_k) \rangle $.
Recall that for four graviton scattering amplitude,
there exist two cases of $\langle i, i+2 \rangle$ where $i=1,2$ 
which do not exist in the correlator of stress-energy tensor.
As we have seen in (\ref{res5}), (\ref{res6}), and 
(\ref{res7}) for four-point function, 
we should consider the following factors for five-point 
function:
\bea
\left(\frac{\pi_1^a \pi_{3a}}{\pi_1^1 \pi_3^1}\right)^{-1},
\quad
\left( \frac{\pi_2^a \pi_{4a}}{\pi_2^1 \pi_4^1} \right)^{-1},
\quad
\left( \frac{\pi_3^a \pi_{5a}}{\pi_3^1 \pi_5^1} \right)^{-1},
\quad
\left( \frac{\pi_4^a \pi_{1a}}{\pi_4^1 \pi_1^1} \right)^{-1},
\quad
\left( \frac{\pi_5^a \pi_{2a}}{\pi_5^1 \pi_2^1} \right)^{-1}
\nonu
\eea
as a prefactor of 
a vertex operator at each external particle. 
Of course, we can consider part of these
and the remaining factors can be obtained by permutations
between particles, as we have seen in four-particle 
scattering amplitude. 
Part of subamplitudes contain the factors
\bea
\frac{\langle i, \zeta \rangle^8}{\langle j, \zeta \rangle^2 
\langle k, \zeta \rangle^2 
\langle l, \zeta \rangle^2  \langle m, \zeta \rangle^2  },
\qquad
\frac{\langle i, \zeta \rangle^6}{\langle j, \zeta \rangle^2 
\langle k, \zeta \rangle^2 
\langle l, \zeta \rangle^2},
\qquad
\frac{\langle i, \zeta \rangle^4 \langle j, \zeta \rangle^2}
{\langle k, \zeta \rangle^2 
\langle l, \zeta \rangle^2 
\langle m, \zeta \rangle^2  }
\nonu 
\eea
where $i,j,k,l,m=1, \cdots, 5$, 
compared with four-point amplitude. 
They are homogeneous of degree zero in sum of each $\la_i$ although
each $\la_i$ has different degrees in each case.
We expect, by counting the independent subamplitudes with correct
multiplicities which will be very tedious, 
that we can write down the MHV tree amplitude for
scattering of five gravitons  will be coincident with 
the result of \cite{BW} eventually.

$\bullet$ Five-point amplitude for one graviton and four gluons 

As we have done in (\ref{tjjjcor}) before, 
the expression for 
the correlator $\langle T(z_1) \prod_{k=2}^{5} J^{r_k} (z_k)  
\rangle$ can be obtained from the basic OPEs (\ref{jjope}) 
and (\ref{tjope}). 
One can think of one of the terms in this correlator:
The factor $\frac{1}{z_{12}^2}$
comes from the contraction between $z_1$ and $z_2$ and there is a
four-point correlator 
$\langle \prod_{k=2}^{5} J^{r_k} (z_k)  
\rangle$ whose explicit form was given in the footnote 
\ref{foot1} already.
The relevant term in the five-point correlator 
will contain $\frac{1}{z_{12}^2 
z_{23} z_{34} z_{45} z_{53}}$.
One expects that the subamplitude for this case looks like
$
(-i) 
 \left( \prod_{k=2}^{5} \frac{1}{\langle k, k+1\rangle} 
\right)
\langle r, s\rangle^4
\frac{[1,2]\langle 2, \zeta \rangle^2 }
{\langle 1,2 \rangle \langle 1,
\zeta \rangle^2}$, as we already mentioned in the last paragraph 
after (\ref{restjjj}). 
Together with other three cases, the full amplitude can be
written similar to (\ref{restjjj}) without any difficulty.

$\bullet$ Five-point amplitude for two gravitons and three gluons 

One can see the term $\frac{1}{z_{12}^2} 
\langle T(z_2) \prod_{k=3}^{5} J^{r_k} (z_k)  
\rangle
$ by contracting of two stress-energy tensor, in the correlator 
$\langle T(z_1) T(z_2) \prod_{k=3}^{5} J^{r_k} (z_k)  
\rangle$. 
We have already seen the four-point function in (\ref{tjjjcor}).
Then it is easy to check that 
the subamplitude for this case corresponds to 
$\frac{1}{z_{12}^2 z_{23}^2 z_{34} z_{45} z_{53}}$. 
Clearly, the factor $\frac{1}{z_{34} z_{45} z_{53}}$ 
will enter into 
the MHV tree amplitude for three gluons.
We will obtain the final expression of this amplitude, 
as mentioned in (\ref{final}).

$\bullet$ Five-point amplitude for three gravitons and two gluons 

Since the contraction $z_1$ and $z_2$ gives 
a term $\frac{1}{z_{12}^2} T(z_2)$, this five-point correlator
contains $\frac{1}{z_{12}^2} \langle T(z_2) T(z_3) 
J^{r_4}(z_4) J^{r_5}(z_5) \rangle$. According to (\ref{ttjjcor}),
one can read off a term like 
$\frac{1}{z_{23}^2 z_{34}^2 z_{45}^2}$ from those four-point 
function. Therefore, one can see  
the total factor 
$\frac{1}{z_{12}^2 z_{23}^2 z_{34}^2 z_{45}^2}$ in the correlator.
The factor $1/z_{45}^2$ will provide the MHV tree amplitude for
two gluons at $z_4$ and $z_5$ and the others with appropriate 
prefactors will give us other factor with gravitons.

$\bullet$ Five-point amplitude for four gravitons and one gluon 

Since the correlator between stress-energy tensor and the current
vanishes, the five-point correlator 
$\langle  \left(\prod_{k=1}^{4} T(z_k) \right)   J^{r_5} (z_5)  
\rangle=0$. There is no contribution in the amplitude.

In this way, one can construct 
higher $N$-point MHV tree amplitudes($N\geq 5$) 
which will coincide with 
the result of \cite{BW}. 

One can think of conformal dimension 3 operator, 
represented by third order Casimir operator of the underlying 
representation of Lie algebra $SU(N)$, taking the form
$W(z) = d_{r_1 r_2 r_3} J^{r_1} J^{r_2} J^{r_3}(z)$ \cite{BS}:
triple product of current. 
Here $d_{r_1 r_2 r_3}= \mbox{Tr}(\{T^{r_1}, T^{r_2} \}T^{r_3} )$ 
is a completely symmetric traceless tensor and its indices are
contracted with  the color indices of the current. 
Then it is easy to see this $W(z)$ has a conformal dimension 3
under the stress-energy tensor $T(z)$.
The four-point function of $W(z)$ \cite{Zamol} can be obtained
from their basic OPE's $W(z_1)W(z_2)$ and 
$T(z_1) W(z_2)$ and is 
given by \footnote{
It can be checked that the two-point and three-point functions 
are given by 
$\langle W(z_1) W(z_2) \rangle =\frac{c/3}{z_{12}^6}$ and
$\langle W(z_1) W(z_2) W(z_3) \rangle =0$.}
$
\langle \prod_{k=1}^{4} W(z_k)  \rangle =
G(x) z_{13}^{-6} z_{24}^{-6}
$
where $G(x) = 
\cdots + 2c \left[ \frac{1}{x^3} + 
\frac{1}{(1-x)^3}\right] + \cdots$ and $x$ is defined as 
(\ref{ttttcor}). 
The complete expression for four-point function is 
already in \cite{Zamol}.
Then it is easy to see that 
the four-point function contains a factor 
$\frac{1}{z_{12}^3z_{23}^3z_{34}^3z_{41}^3}$ as well as other 
singular terms. Recall that $W(z)$ has a conformal dimension 3.
This shares common feature with the four-point functions of 
dimension 1(current) in the footnote 1 and dimension 
2(stress-energy tensor) (\ref{ttttcortrue}).
It would be interesting to see how this dimension 3 operator 
and its correlators occur in the scattering amplitudes 
in context of the topological B-model.

\vspace{1cm}
\centerline{\bf Acknowledgments}
\indent

I would like to thank N. Berkovits, F. Cachazo, C. Hofman, 
P. Svrcek and 
E. Witten for discussions. 
In particular, I thank P. Svrcek for intensive discussions
and collaboration at early stage.
This research was supported by a grant in aid from the 
Monell Foundation through Institute for Advanced Study, 
and by Korea Research Foundation Grant (KRF-2002-015-CS0006).

\end{document}